\documentclass{aa}  
\usepackage{natbib}
\bibpunct{(}{)}{;}{a}{}{,} 
\usepackage[pdfpagelabels]{hyperref}
\usepackage{txfonts}
\usepackage{comment}
\usepackage{enumitem}
\usepackage{booktabs}
\usepackage{color}

\newcommand{\thethree}{{\textsc The300}}
\newcommand{\theth}{{\textsc The Three Hundred }}

\begin{document}

\title{Predicting Halo Formation Time Using Machine Learning}


\author{Atulit Srivastava
          \inst{1,2},
          Weiguang Cui\inst{1,2},  
          Daniel de Andres\inst{1,2}
          \and
          Jesse B. Golden-Marx\inst{3}
          \and Elena Rasia\inst{4,5,6}
          \and Ying Zu \inst{7, 8, 9}
          }
\authorrunning{Srivastava et al.}
\institute{Departamento de Física Teórica, M-8, Universidad Autónoma de Madrid, Cantoblanco E-28049, Madrid, Spain 
\and
Centro de Investigación Avanzada en Física Fundamental (CIAFF), Universidad Autónoma de Madrid, Cantoblanco E-28049, Madrid, Spain \and 
School of Physics and Astronomy, University of Nottingham, Nottingham,
NG7 2RD, UK
\and
INAF, Osservatorio Astronomico di Trieste, via Tiepolo 11, I-34143 Trieste, Italy
    \and
    IFPU, Institute for Fundamental Physics of the Universe, Via Beirut 2, I-34014 Trieste, Italy
    \and
    Department of Physics; University of Michigan, 450~Church~St, Ann Arbor, MI~48109, USA
    \and
Tsung-Dao Lee Institute \& Department of Astronomy, School of Physics and Astronomy, Shanghai Jiao Tong University, Shanghai 200240, China
\and
Shanghai Key Laboratory for Particle Physics and Cosmology, Shanghai Jiao Tong University,
Shanghai 200240, People’s Republic of China
\and
Key Laboratory for Particle Physics, Astrophysics and Cosmology, Ministry of Education, Shanghai Jiao Tong University, Shanghai 200240, People’s Republic of
China
}

   \date{November 26, 2024}

 
  \abstract
   {
   The formation time of dark-matter halos quantifies their mass assembly history, and as such, it directly impacts the structural and dynamical properties of galaxies within and even influences their galaxy evolution. Despite its importance, halo formation time is not directly observable, necessitating the use of indirect observational proxies—often based on star formation history or galaxy spatial distributions. Recent advancements in machine learning allow for a more comprehensive analysis of galaxy and halo properties, making it possible to develop models for more accurate prediction of halo formation times.
    }
   {This study aims to investigate a machine learning-based approach to predict halo formation time—defined as the epoch when a halo accretes half of its current mass—using both halo and baryonic properties derived from cosmological simulations. By incorporating properties associated with the brightest cluster galaxy located at the cluster center, its associated intracluster light component and satellite galaxies, we aim to surpass these analytical predictions, improve prediction accuracy and identify key properties that can provide the best proxy for the halo assembly history.}
   {Using \theth\ cosmological hydrodynamical simulations, we train Random Forest (RF) and Convolutional Neural Network (CNN) models. The random forest models are trained using a range of dark matter halo and baryonic properties, including halo mass, concentration, stellar and gas masses, and properties of the brightest cluster galaxy and intracluster light within different radial apertures, while CNNs are trained on two-dimensional radial property maps generated by binning particles as a function of radius. Based on these results, we also construct simple linear models that incrementally incorporate observationally accessible features, optimizing for minimal bias and scatter for predicting the halo formation time.
}
   {Our RF models demonstrated median biases between 4\% and 9\%
   with relative error standard deviations around 20\%
   in the prediction of the halo formation time. CNN models trained on two-dimensional property maps, further reduced the median bias to $\lesssim 4\%$,
   though with a higher scatter than the random forest models. With our simple linear models, one can easily predict the halo formation time with only a limited few observables with the bias and scatter compatible with RF results. Lastly, we also show that the traditional relations between halo formation time and halo mass or concentration are well preserved with our predicted values.
   }
  {}
   \maketitle
%
\section{Introduction}\label{section:1}
Dark matter halos are virialized structures within the cosmological constant-dominated cold dark matter cosmological model, $\Lambda$CDM.
They form hierarchically, starting from initial density peaks generated by the early universe perturbations. The matter then collapses into these peaks due to gravitational instability, and over time, these smaller halos merge to form larger ones.
Within halos, luminous galaxies are formed when the baryonic gas cools and condenses within the deep potential wells laid out by these dark matter halos  \citep[e.g.][]{10.1093/mnras/183.3.341, mo2010galaxy}. It is interesting to note that not all massive halos at high redshift are also the most massive ones at $z=0$ as a result of the different halo formation histories (Onions et al. in prep.). 
Hence, to obtain a deeper understanding of the formation and evolution of galaxies, which is closely tied to the coevolution of their host dark matter halos, one needs to thoroughly understand the formation history of the halos. The most important quantity of a halo is its mass, as such its formation history is typically quantified by its mass growth -- using a single-parameter measure called halo formation time \citep[see e.g.][and citations thereafter]{Wechsler2002,gao2005age}, defined as the epoch when the dark matter halo acquires half of its final mass.
This is one of many approaches to quantify halo assembly history, with the choice of definition often depending on the specific goals of the study \citep[see][for alternative definitions and their distinctions]{li2008halo, Tojeiro2017}. The definition selected in this study, based on the hierarchical nature of halo formation, is the most widely used one.

Halo formation time, hereafter $t_{1/2}$, as the second important quantity of the dark matter halo, influences many other galaxy/halo properties and statistics. $t_{1/2}$ directly connects with the halo's concentration parameter and dynamical states \citep[e.g.][]{power2012dynamical,wong2012dark,mostoghiu2019three, Chen2020}. Given that the halo mass growth links to its merger history, it is not surprising to see early-formed halos tend to be dynamically relaxed with a higher concentration. 
At the same time, these other two properties, halo dynamical state, and concentration, which are easier to measure, have been linked to 
a variety of other halo properties, such as halo mass estimation bias \citep[e.g.][for hydrostatic equilibrium method]{gianfagna2023study}, protohalo size\citep{Wang2024}, backsplash galaxies \citep[e.g.][]{haggar2020thethreehundred}, the $M_{\text{sub}} - V_{\text{circ}}$ relation \citep{srivastava2024three}, the connectivity \citep{santoni2024three}, X-ray brightness \citep{ragagnin2022simulation,bartalucci2023chex}, and more. Lastly, $t_{1/2}$  also drives the so-called halo formation bias effect \citep{gao2005age, Wechsler2006, li2008halo, croton2007halo}, which affects the clustering statistics between dark matter halo and the underlying large-scale matter distribution -- halo bias\citep[see][for a detailed review]{desjacques2018large}. Neglecting this dependence can lead to incorrect inferences about galaxy clustering and inaccurate constraints on cosmological parameters.


Besides the halo properties, $t_{1/2}$ can also influence the galaxy properties within the halos, such as the central galaxy stellar-to-halo mass (SMHM) relation \citep[e.g.]{matthee2016origin,zehavi2018impact}. Recently, \cite{Cui2021} found that $t_{1/2}$ is the intrinsic driver for the SMHM colour bimodality, condition verified in galaxy groups \citep[e.g.][]{cui2024hyenas} as well as galaxy clusters \citep[e.g.][]{zu2021does}.
%
Moreover, at cluster mass scale, the evolutionary history of the host halo also impacts on the central brightest galaxy (BCG) \footnote{BCGs refer to the Brightest Central Galaxies or the Brightest Cluster Galaxies.}
and the surrounding, faint intra-cluster light (ICL) \citep{zibetti2005intergalactic, jee2010tracing, contini2020mass, contini2021brightest}. Because the ICL is mainly formed through merger events \citep{contini2014formation,burke2015coevolution,jimenez2018unveiling}, it is expected that earlier-formed dark matter halos correspond to earlier-formed ICL \citep{contini2023connection}. As such, relaxed clusters, which are predominantly early-formed, have a significantly higher mass fraction of ICL and BCG compared to unrelaxed clusters, which are mostly late-formed \citep{contini2014formation, cui2014characterizing,chun2023formation}. Furthermore, \cite{Alonso2020} and \cite{contreras2024characterising} showed that the ICL profile scales with the dark matter density profile, which tightens the ICL connection to its host halo. 
Therefore, accurate estimation of $t_{1/2}$ is pivotal for further deepening our understanding of the connection between the host halo and its stellar components
\citep{wechsler2018connection}.

Despite its importance, $t_{1/2}$ is not a quantity that can be directly derived from any observation, therefore a proxy is needed.
For instance, the connection between halo formation time and cluster dynamical state/concentration can be used for $t_{1/2}$ estimation.
However, the halo dynamical state is also an ill-defined quantity in both observations and simulations, since each of the many parameters used refers to a peculiar condition of relaxed/disturbed systems, see \cite[e.g.][for research along this line]{rasia.etal.2013,cui2017dynamical,haggar2020thethreehundred,de2021three, capalbo2021three, zhang2022three}. Furthermore, both the observation-to-theoretical dynamical parameter and the dynamical parameter to $t_{1/2}$ links show a large scatter \citep{darragh-ford.etal.2023}. Because of measuring the total halo density profile (see e.g. \cite{Umetsu2016} for lensing method, \cite{Alonso2020,contreras2024characterising}) for using Intra-cluster light, \cite{ferragamo2023three} for ML method) is not easy and the connection between concentration and $t_{1/2}$ is not very strong (see this work), using concentration to predict $t_{1/2}$ seems not very promising. 
From the observational side, $t_{1/2}$ has been linked to the following galaxy or member galaxy properties: the spatial distribution of member galaxies within groups or clusters \citep{miyatake2016evidence, more2016detection, zu2017level}; the central galaxies properties within these groups or clusters \citep{wang2013detection,lin2016detecting, lim2016observational}; galaxy colour or SFR \citep{yang2006observational, wang2013detection, watson2015predicting}, and the magnitude gap between the BCG and either the second (M12) or fourth brightest galaxy (M14) \citep{golden2018impact, vitorelli2018mass, golden2019impact, golden2022observed, golden2024hierarchical}. However, \citet{lin2016detecting} ruled out the connection between $t_{1/2}$ and galaxy SFR/colour and \citet{zu2017level, sunayama2019measurements} have shown that using member galaxies as a proxy for formation time is susceptible to optical projection effects, which can lead to false detection. By studying the central BCG properties and profiles, \cite{golden2022observed, golden2024hierarchical} suggested that central galaxy growth is less related to its host halo growth, while the ICL is more strongly related to both the $t_{1/2}$ and  $m_{\text{gap}}$.
Given the contrasting results, which are plagued by systematic effects, and uncertainty regarding which galaxy property reliably indicates the $t_{1/2}$, a detailed investigation is imperative to develop a framework that can provide a reliable estimation of $t_{1/2}$.

A supervised machine learning (ML) framework offers an effective approach to this problem by incorporating various galaxy properties as features in a model that, once trained, can accurately predict different unknown properties. This approach leverages ML's ability to learn complex relationships within multidimensional datasets.
ML has been successfully applied to a wide range of astrophysical problems, including the construction of cosmological hydrodynamical simulations \citep{villaescusa2021camels}, predicting baryonic properties (star and gas) from dark matter halo properties \citep{de2023machine}, obtaining cluster mass estimates \citep{armitage2019application,de2022deep,ferragamo2023three,de2024three}, studying galaxy merger events \citep{contreras2023galaxy,arendt2024identifying} and modeling the galaxy-halo connection \citep{lovell2022machine, wadekar2020modeling}, among others. More recently, deep learning techniques have also been employed in various studies, where convolutional neural networks were trained using the observable properties of simulated galaxies \citep{sullivan2023learning} or the local formation processes of simulated dark matter halos \citep{lucie2023halo} to predict halo assembly bias parameters. 
To this end, we use The Three Hundred Cosmological hydrodynamical (\thethree) zoom-in simulations of galaxy clusters introduced in \citet{cui2018three} to explore the potential of employing an ML framework to link several galaxy, BCG-ICL properties to the formation time of their dark matter host halos. However, one of the challenges that has remained unresolved is the identification of the BCG and its associated ICL, both observationally and in simulations 
\citep[e.g.,][]{gonzalez2005intracluster, krick2007diffuse, jimenez2016disentangling}.
Observationally, separating the ICL from the BCG is difficult due to the blending of surface brightness and contamination from foreground and background sources
\citep{mihos2005diffuse,zibetti2005intergalactic,rudick2010optical}.
In simulations, even with complete particle information, distinguishing the two components is complicated by the overlapping gravitational potentials of the host cluster and its satellite galaxies \citep[e.g.,]{cui2014characterizing, remus2017co}. Different studies adopt various approaches to define the BCG and ICL, with a commonly used method in simulations being a fixed aperture radius, typically between 30 and 100 kpc \citep[e.g.,]{mccarthy2010case,kravtsov2018stellar,pillepich2018first}.
This method ensures consistency with observational studies that also rely on fixed apertures \citep{stott2010xmm}. Previous analyses of The Three Hundred simulations have followed this approach \citep{contreras2022three}, making it a suitable choice for our study. For other galaxy properties, such as gas metallicity, we acknowledge that these may not yet be easily accessible. However, we are optimistic that future telescopes will provide the necessary measurements.

The paper is organized as follows. In Section \ref{section:2}, we describe the \theth\ simulation project and detail the halo, galaxy, and BCG-ICL properties used to train the ML models. 
In Section \ref{section:3} introduces the random forest regressor algorithm \citep{breiman2001random} that would use all the properties discussed in Section \ref{section:2} and presents an assessment of the $t_{1/2}$ predictions derived from RF models trained on different combinations of halo and galaxy properties derived from the simulation dataset.
In Section \ref{section:4}, we train a convolutional neural network using galaxy stellar and gas properties, along with the BCG+ICL system, to assess the model's performance in predicting 
\(t_{1/2}\). We further benchmark the performance of ML models in Section \ref{section:correlation and EPS}. 
In Section \ref{section:5}, we present simple linear models for estimating \(t_{1/2}\) , which perform comparably to the more complex machine learning models, albeit with a slight increase in the scatter of the predictions.
We conclude in Section \ref{section:6}. 

\section{Simulation}\label{section:2}
We use a sample of galaxy clusters based on the suite of hydrodynamical simulations from \theth\ project (hereafter \thethree, \citealt{cui2018three}). \thethree\ project targets spherical regions centered on the 324 most massive clusters from the original dark-matter-only Multidark simulation (MDPL2, \cite{klypin2016multidark}) with a box size of $1 h^{-1} \text{Gpc}$.
These large zoomed-in regions, each with a radius of $15 h^{-1} \text{Mpc}$, are then re-simulated at a high resolution consistent with the Planck cosmology as the parent dark-matter-only simulation (adopting the cosmological parameters $\Omega_{M}$ = 0.307, $\Omega_{B}$ = 0.048, $\Omega_{\lambda}$ = 0.693, h = 0.678, $\sigma_{8}$ = 0.823, $n_{s}$ = 0.96 \citealt{planck2016cosmological}), incorporating different flavours of baryon physics in the zoomed-in regions. The dark matter and gas-particle masses within the high-resolution region for \thethree\ simulated clusters are $m_{\text{DM}} \approx 12.7 \times 10^8 h^{-1} M_\odot$ and $m_{\text{gas}} \approx 2.36 \times 10^8 h^{-1} M_\odot$, respectively.  Beyond the $15 h^{-1} \text{Mpc}$ region, the rest of the simulation is populated with low-resolution mass particles to simulate large-scale tidal effects in a computationally efficient way compared to the original MDPL2 simulation. This work focuses on the galaxy clusters re-simulated with the Gizmo-Simba baryonic physics model \citep{dave2019simba, cui2022three}. For each cluster simulated in the \thethree\ project, we save 128 snapshot files corresponding to redshifts ranging from $z = 16.98$ to $0$. 

\subsection{Halo  and Galaxy Catalogs}
The catalogs for dark matter host halos and substructures for each cluster region in \thethree\ project simulation are identified using the Amiga Halo Finder (AHF) \citep{knollmann2009ahf}. AHF utilizes the adaptive mesh refinement technique, in which the density field of the simulation is successively divided to locate spherical over-density peaks, considering all dark matter, stars, gas, and black hole particle species. 
The halos are identified as spherical regions with an average density of 200 times the universe's redshift-dependent critical density, $\rho_{\rm crit}(z)$. The mass \(M_{200c}\) of the identified spherical region is given by \(200 \times (4\pi/3) R_{200c}^{3} \  \rho_{crit}(z)\), where \(R_{200c}\) is the radius of the identified spherical region. In this study, we selected a sample of 1,925 dark matter halos with \(M_{200c}\) ranging from \(2.50 \times 10^{13}\) to \(2.64 \times 10^{15} \, h^{-1} \, \text{M}_{\odot}\) from the \(z=0\) snapshots of the \thethree\ simulated cluster regions.
In Table \ref{Table:1}, we provide the AHF-computed halo physical properties at $z=0$ which will be utilized as features within the RF regressor for predicting $t_{1/2}$.
\begin{table*}[h]
    \renewcommand{\arraystretch}{1.5} 
    \centering
    \caption{List of the halo properties at $z=0$ extracted from AHF and utilized within this study for training the RF regressor.}
    \label{Table:1}
    \begin{tabular}{p{3.2cm}p{8.8cm}p{2cm}}
        \toprule
        \textbf{Property} & \textbf{Description} & \textbf{Units} \\
        \midrule
        \textit{$M_{\mathrm{gas}}$} & The gas mass within the halo radius $R_{200c}$. & $h^{-1} M_{\odot}$ \\
        \textit{$M_{*}$} & The stellar mass within the halo radius $R_{200c}$. & $h^{-1} M_{\odot}$ \\
        \textit{$M_{200c}$} & The halo overdensity mass. & $h^{-1} M_{\odot}$ \\
        \textit{$N_{sub}$} & The number of sub-structures within a halo. & - \\
        $\Phi_0$ & The potential corresponding to the integration constant that needs to be evaluated within AHF to remove gravitationally unbounded particles from the identified halos. & (km/sec)$^{2}$ \\
        $R_{200c}$ & The halo overdenisty radius. & $h^{-1}$ kpc \\
        \textit{$V_{max}$} & The maximum circular velocity of the halo which is determined from the rotation curve considering both the dark matter and baryonic particles within the halo. & km/s \\
        \textit{b} & The ratio between the second largest ($b$) and the largest ($a$) eigenvalue of the moment of inertia tensor. & $b/a$ \\
        \textit{c} & The ratio between the smallest ($c$) and the largest eigenvalue of the moment of inertia tensor. & $c/a$ \\
        \textit{$c_{\mathbf{NFW}}$} & The dimensionless concentration parameter which characterizes the Navarro–Frenk–White (NFW, \cite{navarro1997universal}) density profile for a dark matter halo. Here, we employ a simpler version to estimate the halo concentration using the velocity ratio described in \citet{prada2012halo}. & - \\
        \textit{com\_offset} & The distance between the center of mass of the halo and its density peak which is used as an indicator to describe the halo’s dynamical state (see \citet{10.1093/mnras/stw2567, haggar2020thethreehundred} for example). & $h^{-1}$ kpc \\
        \textit{$\lambda$} & Spin measure of the halo based on \citet{bullock2001profiles} definition. & - \\
        \textit{mean $Z_{\mathrm{gas}}$} & Mean gas metallicity. & $Z_{\odot}$ \\
        \textit{mean $Z_{*}$} & Mean stellar metallicity. & $Z_{\odot}$ \\
        \textit{$\sigma_V$} & The 3D velocity dispersion for all the particles inside the halo. & km/s \\
        \textit{$V_{esc}$} & The escape velocity at $R_{200c}$. & km/s \\
        $E_{kin}$& Total kinetic energy of the halo. &$h^{-1}M_{\odot}$(km/s)$^{2}$\\
        $E_{pot}$& Total potential energy of the halo. &$h^{-1}M_{\odot}$(km/s)$^{2}$\\
        $E_{surf}$& Surface pressure at the boundary of the halo due to collisionless particles \citep{shaw2006statistics}. &$h^{-1}M_{\odot}$(km/s)$^{2}$\\
        \bottomrule
    \end{tabular}
\end{table*}

Some galaxy quantities are calculated using the \textsc{CAESAR} catalog, e.g. the galaxy half mass radius and stellar masses. \textsc{CAESAR}\footnote{\url{http://caesar.readthedocs.org/}} is a galaxy finder, which possesses the flexibility to run on halo catalogs generated by different halo finders. The galaxy catalogs we used in this paper are run on the \textsc{AHF} halo catalog. For more details of these galaxy properties and how they are identified, we refer interested readers to \cite{cui2022three}.

\subsection{Merger tree and $t_{1/2}$}
MERGERTREE, a tool provided within the AHF package, is then used to make halo merger trees to trace the evolutionary history of halos with redshifts which involves building links between halos through various simulation snapshots. 
The mechanism for identifying the main progenitor of each halo, starting from redshift zero, involves back-tracing that halo to the previous snapshot. This is done by calculating the merit function of that halo with respect to each halo present within the previous snapshot and selecting the one that maximizes the merit function.
The merit function used to generate the merger tree is defined as \(N_{ab}^{2} / N_{a} N_{b}\), where \(N_{a}\) is the number of particles within the halo for which we want to identify the main progenitor halo in the previous snapshot, \(N_{b}\) is the number of particles within the possible main progenitor halo of the previous snapshot, and \(N_{ab}\) is the number of particles shared between the two halos. Furthermore, the merger-tree generator can skip over snapshots where a suitable progenitor halo cannot be identified,  ensuring a continuous tracking of the evolution of the halo across the available snapshots. For more details regarding the MERGERTREE one can refer to \citet{srisawat2013sussing}. 

$t_{1/2}$ for these selected samples of halos is then estimated in terms of redshift by tracing the most massive progenitor at each redshift on the main branch of the merger tree and arranging them chronologically, providing us with the halos' mass accretion history.
\begin{figure}[h]
    \centering
    \includegraphics[scale=0.4]{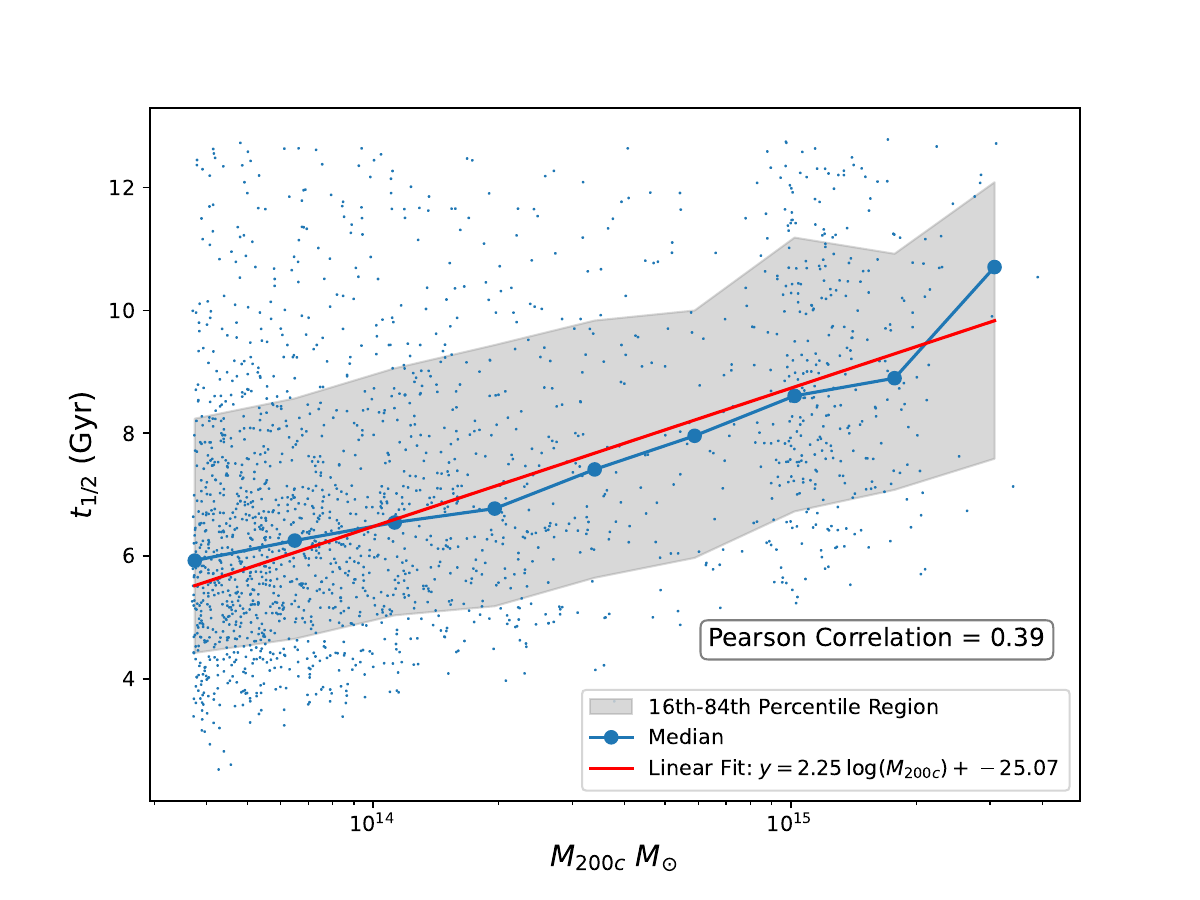}
    \caption{Scatter plot depicting the relationship between the logarithmic halo mass at $z=0$ and $t_{1/2}$ in Gyr for the selected sample of halos from \thethree. The red line represents the best-fit line that has been obtained by fitting a simple linear relation between the halo mass and the $t_{1/2}$.}
    \label{fig:1}
\end{figure}
We converted all estimated halo formation times values from redshift into giga-year ($t_{1/2}$ Gyr) -- the universe age, hence all mentions of halo formation time in the paper will be in Gyr.
These $t_{1/2}$ values of the selected sample will serve as the target variable that the ML model will try to predict.
In Figure \ref{fig:1}, we present a scatter plot between the halo mass, $(M_{200c})$, and $t_{1/2}$, illustrating the positive correlation (as indicated by the Pearson coefficient value of 0.39 in the plot) between the final assembled halo mass at $z=0$ and the $t_{1/2}$ for our selected halos sample. The blue line with the scatter points depicts the median trend, which is obtained by computing the median 
$t_{1/2}$ in the logarithmic mass bins. The red line represents the linear relation, which is obtained by fitting a linear model between the center of the mass bins and $t_{1/2}$. The plot suggests that massive halos tend to be late-forming compared to the low-mass halos, thus clearly manifesting the hierarchical assembly growth of halos as predicted by \(\Lambda\)CDM cosmology. However, it is worth noting that there is a huge scatter in this plot, i.e. there are massive halos that can be formed very early, only $\sim 4, 5$ Gyr of the Universe, and vice versa.
In the following subsection, we will explore the properties derived from the simulation of the embedded BCG+ICL system, which will be utilized for the RF regressor.

\subsection{BCG and ICL properties}
As discussed in the introduction, the $t_{1/2}$ can be connected to the surrounding BCG area, and, thus, to the ICL. However, a clear separation of the BCG and ICL systems within the halos in both simulations and observations is non-trivial, given that the two components are highly blended. However, a common approach is to employ a fixed aperture radius and consider all the particles within that sphere to identify the BCG region in simulation-based studies or a fixed magnitude limit in observation-based studies. For example, \citet{mccarthy2010case} used a 30 $h^{-1}$ kpc radius aperture, \citet{contreras2024characterising} employed a 50 $h^{-1}$ kpc radius, and \citet{contreras2022three} used three different aperture radii of 30, 50, and 70 $h^{-1}$ kpc to define the BCG region. While observers can adopt surface brightness cuts \citep[e.g.][and references therein]{zibetti2005intergalactic} to separate BCG and ICL. We refer interesting readers to \cite{cui2014characterizing, Presotto2014, brough2024preparing} for comparisons between different separation methods. Following \cite{contreras2022three}, we simply employ three different 3D aperture radii to define the BCG in our halo dataset: 30, 50, and 100 $h^{-1}$ kpc. The larger radii 100 $h^{-1}$ kpc 
includes the external wings of the BCG. 
We select all the gas, stars, black holes, and dark matter particles within the spheres defined by these three different aperture radii to define the BCG region. After defining the BCG, we identify the ICL for our simulated halos by selecting all the star particles that are bound to the halo potential but are not part of either the BCG or any other substructures within the halos. This ensures that the ICL represents stars that are not gravitationally bound to individual galaxy members or merged groups but are instead spread throughout the cluster. Indeed, in general, the ICL covers large portions of the cluster. 
It is also important to note that ICL used in \autoref{Table:2} refers to halos where the BCG was determined using a 50 kpc aperture radius 3D cut. 
 
For this, we use the AHF\textunderscore particles file generated by the AHF, which includes all the IDs of the particles that belong to each halo. After having the BCG and ICL regions identified, we compute their physical properties in order to utilize them within the RF regressor.
Using the simulated halo dataset, we first computed the properties corresponding only to the combined BCG and ICL region identified using the 50 \(h^{-1}\) kpc aperture 3D cut which is listed in Table \ref{Table:2}. These properties will be considered in our RF model as features along with the halo properties computed using AHF (i.e. Table \ref{Table:1}).    
\begin{table*}[h]
    \renewcommand{\arraystretch}{1.5} 
    \centering
    \caption{List of properties for the BCG and ICL region identified using the 50 $h^{-1}$ kpc aperture radius for training the RF regressor.}
    \label{Table:2}
    \begin{tabular}{p{3.2cm}p{8.8cm}p{2cm}}
        \toprule
        \textbf{Property} & \textbf{Description} & \textbf{Units} \\
        \midrule
        \textit{\large $\frac{M_{BCG}}{ICL}$} & The ratio between the total BCG mass computed within the 50 \(h^{-1}\) kpc aperture 3D cut and the ICL stellar mass. & - \\
        \textit{$M_{bh}$} & The total mass of black hole particles at the center of the BCG.. & $h^{-1} M_{\odot}$ \\
        \textit{\large $ M_{12} = \log_{10}\left(\frac{M_{BCG}}{M_{\text{sub}2}}\right)$} & The logarithmic mass difference between the total BCG mass computed within the 50 \(h^{-1}\) kpc aperture 3D cut and the most massive substructure. This is akin to the magnitude gap definition $M12$. 
        & - \\
        \textit{\large $M_{14} = \log_{10}\left(\frac{M_{BCG}}{M_{\text{sub}4}}\right)$} & The logarithmic mass difference between the total BCG mass computed within the 50 \(h^{-1}\) kpc aperture 3D cut and the fourth massive substructure. This is akin to the magnitude gap definition $M14$. & - \\
        \textit{\large$\frac{M_{BCG}}{M_{sat}}$} & The ratio between the total BCG stellar mass computed within the 50 \(h^{-1}\) kpc aperture and the total satellite stellar mass of the halo.  & - \\
        \textit{dist} &  The three-dimensional distance between the center of the BCG and the center of mass calculated from all the satellite galaxies in terms of their stellar masses. & $h^{-1}$ kpc \\
        \bottomrule
    \end{tabular}
\end{table*}
In Table \ref{Table:3}, we enumerate all the stellar and gas properties for the BCG defined using three different aperture radii (30, 50, and 100) \(h^{-1}\) kpc for our selected halo dataset from \thethree\ simulated clusters. 
\begin{table*}[h]
\renewcommand{\arraystretch}{1.5} 
\centering
\caption{List of baryonic properties calculated for the BCG defined using three different aperture radii (30, 50, and 100) \(h^{-1}\) kpc, which will be used in this study to train the RF regressor.}
\label{Table:3}
\begin{tabular}{p{3.2cm}p{8.8cm}p{2cm}}
\toprule
\textbf{Property} & \textbf{Description} & \textbf{Units} \\
\midrule
\textit{$t_{30}, t_{50}, t_{100}$} & Mass-weighted stellar age within BCG apertures of $30, 50,$ and $100$ kpc, respectively & Gyr \\
\textit{$f_{gas,30}, f_{gas,50}, f_{gas,100}$} & Mass of the gas particles within BCG apertures of $30, 50,$ and $100$ kpc relative to the halo mass, respectively & - \\
\textit{$f_{*,30}, f_{*,50}, f_{*,100}$} & Mass of the star particles within BCG apertures of $30, 50,$ and $100$ kpc relative to the halo mass, respectively & - \\
\textit{$Z_{gas,30}, Z_{gas,50}, Z_{gas,100}$}& Mass-weighted gas metallicity within BCG apertures of $30, 50,$ and $100$ kpc, respectively & $Z_{\odot}$ \\
\textit{$Z_{*,30}, Z_{*,50}, Z_{*,100}$}& Mass-weighted stellar metallicity within BCG apertures of $30, 50,$ and $100$ kpc, respectively & $Z_{\odot}$ \\
\bottomrule
\end{tabular}
\end{table*}

\begin{figure}[h]
    \centering
    \includegraphics[scale=0.4]{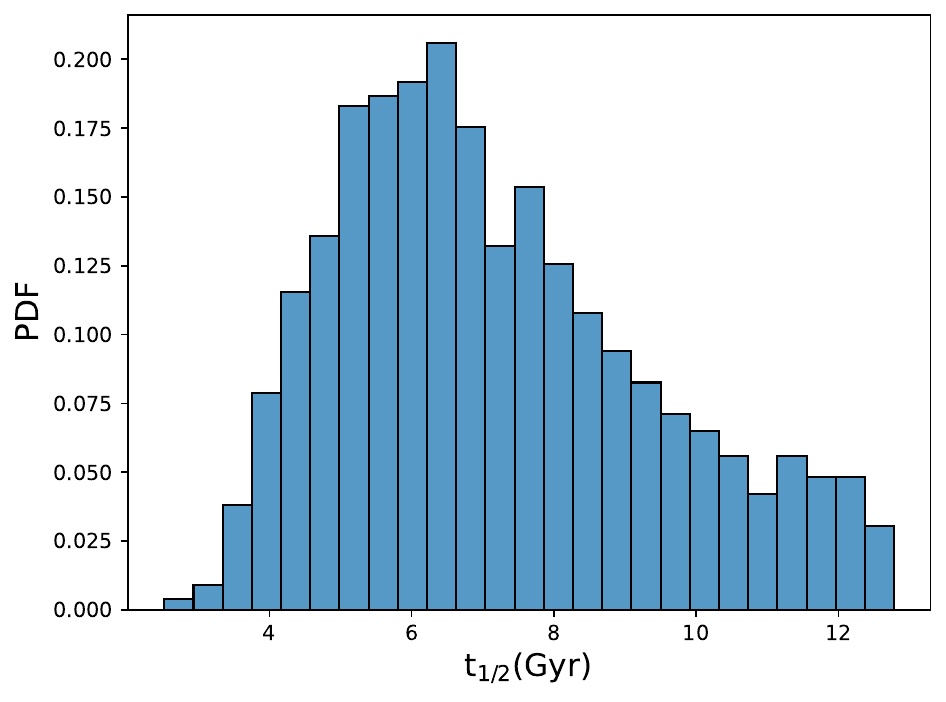}
    \includegraphics[scale=0.4]{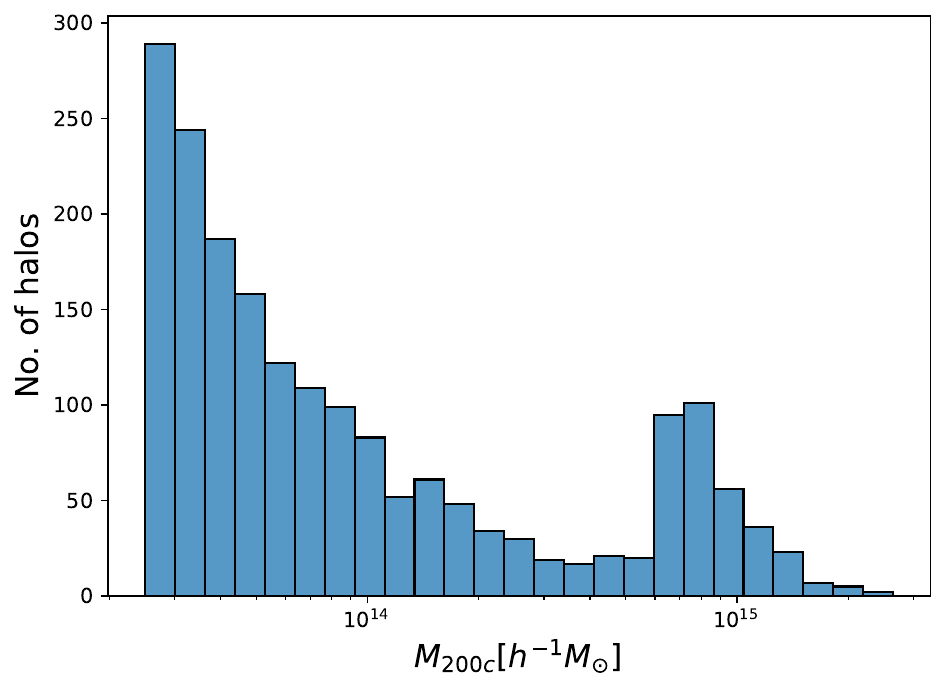}
    \caption{The upper panel depicts the PDF of the $t_{1/2}$
    values denoted by $t_{1/2}$ and the lower panel gives the histogram for the mass distribution for our selected halos sample used in this study.}
    \label{fig:2}
\end{figure}

After obtaining all the properties associated with the halos and their corresponding BCG regions, we proceeded with data preprocessing. We found four halos with a black hole mass $M_{bh}$  
equal to zero, so we removed these halos from our dataset, including all associated properties from Table \ref{Table:1}, Table \ref{Table:2}, and Table \ref{Table:3}. We also found one halo with an $M_{14}$ 
unrealistically high due to fewer numbers of satellites, so we removed that halo from our analysis. Finally, we noted that $f_{gas,30}$ values (defined in Table \ref{Table:3}) were not well defined for two halos, so we excluded them from our analysis as well. 
Our final halo dataset consists of $1,918$ halos for which the probability density function (PDF) of their computed $t_{1/2}$ and histogram of mass distribution is given in Figure \ref{fig:2}. It is clear that neither the $t_{1/2}$ nor the halo mass is uniformly distributed.
As presented in previous ML application studies \citep[e.g.][]{de2022deep,de2024three}, the non-uniform property distribution will bias the ML models toward the results with the highest statistics and thus those studies aimed at creating uniformly distributed datasets to avoid this source of bias.
In our study, we implemented a weight to all the following ML training based on the distribution of the $t_{1/2}$ in \autoref{fig:2}. While training the various machine learning models, to address the issue of undersampling in the target variable (i.e., very few observations for very early- and late-forming halos, as depicted in the upper panel of Figure \ref{fig:2}), we assigned weights to the halos corresponding to the training dataset.
We adopted the inverse of the PDF of $t_{1/2}$ as the weight for each halo with formation time $t_{1/2}$ in the training set and normalized them so that the sum of all weights is unity.
A bump is also noticeable in the mass distribution (Figure \ref{fig:2}, lower panel), which arises due to the completeness limit of our sample (see Appendix \cite{cui2018three}). Our simulations are zoom-in re-simulations of 324 massive regions identified in the MDPL2 dark matter–only simulation. These regions span a large volume that includes not only the central clusters but also many additional halos. While the halo mass function from the full cosmological volume shows the expected smooth curve, our re-simulated sample becomes incomplete below a certain mass threshold. This incompleteness manifests as the observed bump, marking the mass limit above which our sample reliably includes halos from the full volume. 

In the next section, we will explain our analysis that uses the properties described in Table \ref{Table:1}, Table \ref{Table:2}, and Table \ref{Table:3} to train the random forest regressor to predict the $t_{1/2}$.

\section{Random Forest Regression}\label{section:3}
The random forest (RF) algorithm \citep{breiman2001random} is an ensemble-based learning method that combines multiple base learners or regressors to handle multi-dimensional datasets, offering advantages in solving complex problems and improving model performance. The base learners in the RF algorithm are decision trees, each trained independently from the others during the training procedure.
The algorithm utilizes bootstrap sampling with replacement, or bagging, of the original training dataset, which has $N$ rows and $M$ columns. Each tree is trained on a different bootstrap sample. In bagging, each sample consists of the same number of $N$ rows as the original dataset, with some entries potentially repeated and others omitted. Furthermore, each decision tree receives a different set of $m$ features or columns ($m < M$) from the original dataset, ensuring that the features are not correlated. 
Each decision tree is constructed using its $m$ subset of features, where the node split in the tree is determined based on the features within the bootstrapped sample extracted from the original training dataset. The node feature and its splitting threshold are decided according to the mean squared error minimization criterion.  
This process is then executed recursively until we reach a termination condition, such as the maximum depth of the tree or the minimum number of samples per leaf. Once all the decision trees are trained, their predictions are aggregated by averaging the outputs for regression tasks. This aggregation ensures that the final result is robust and not prone to overfitting. For our study, we used the \texttt{scikit-learn} implementation of the random forest regressor \citep{pedregosa2011scikit}. The algorithm involves several key hyperparameters that need to be tuned to achieve optimal model performance. The important hyperparameters considered in this study for tuning the RF model are listed below, along with their definitions:
\begin{itemize}
    \item \textit{n\_estimators}: This parameter determines the number of decision trees used in the random forest. We considered various values for this parameter.
    
    \item \textit{max\_features}: This parameter defines the number of subset features, denoted as $m$ ($<M$-the total number of available features), typically on the order of $\sqrt{M}$ or $\log_{2} M$.
    
    \item \textit{max\_depth}: We explored the maximum depth of the decision trees in the RF model.
    
    \item \textit{min\_samples\_leaf}: The minimum number of samples required to be at a leaf node of the decision trees.
\end{itemize}
The \texttt{GridSearchCV} method within the \texttt{scikit-learn} package is used to tune these hyperparameter values of the RF model. This method works by creating a high-dimensional grid using the range of values for hyperparameters discussed earlier. \texttt{GridSearchCV} then employs K-fold cross-validation, where the training dataset is divided into K equal folds. The K-1 folds are utilized for training, while the remaining fold is reserved for cross-validation. This process is repeated K times, ensuring that each fold serves as a cross-validation dataset. For each coordinate representing a combination of hyperparameters from the high-dimensional grid, \texttt{GridSearchCV} utilizes the training data and evaluates performance using a specified metric (such as accuracy, mean squared error, etc.) through the K-fold cross-validation process described earlier. After evaluating the metric for all combinations in the high-dimensional grid, \texttt{GridSearchCV} selects the best combination of parameters that yields the most optimal performance. In the next subsection, we discuss the training setup we utilized. 
\subsection{RF Training Setup}\label{subsection:3.1}
From our sample of halos, we chose 1,630 halos (85\% of the total) for our training set and 288 halos (15\%) for our the test set. This train-test split of the dataset remains fixed throughout to maintain the consistency of the analysis. For the \texttt{GridSearchCV}, the set of values for the hyperparameters is as follows:
\begin{itemize}
    \item \textit{n\_estimators}: [50, 100, 200, 250, 300, 350, 400, 500],
    \item \textit{max\_features}: [$\sqrt{M}$ or $\log_{2} M$],
    \item \textit{max\_depth}: [6, 7, 8, 10, 13, 15, 17, 19],
    \item \textit{min\_samples\_leaf}: [20, 50, 70, 95, 110, 130].
\end{itemize}
We used \texttt{GridSearchCV} with 5-fold cross-validation with the scoring parameter set to \textit{neg\_mean\_squared\_error}. 
The hyperparameter tuning algorithms within the Scikit-Learn package are conventionally implemented to maximize the scoring function; however, for model performance metrics like mean squared error, the objective is to minimize them, as lower values indicate better performance. Therefore, maximizing the \textit{neg\_mean\_squared\_error} is equivalent to minimizing the mean squared error. The hyperparameter grid and its corresponding search space are utilized to tune the hyperparameters for each RF model individually which means that each model is optimized separately, resulting in its own set of hyperparameters based on its specific training and validation performance. 

Using the aforementioned setup, we applied our halo sample dataset (i.e., Tables \ref{Table:1}, \ref{Table:2}, and \ref{Table:3}) to train the RF regressor algorithm as follows:
\begin{description}
\item[\textbf{Model 1}]: RF trained using the properties listed in Table \ref{Table:1} (AHF-computed simulation properties).
\item[\textbf{Model 2}]: RF trained using the properties listed in Table \ref{Table:2} (properties connecting the BCG+ICL region with the halos).
\item[\textbf{Model 3}]: RF trained using the properties listed in Table \ref{Table:3} (stellar and gas properties defined within different 3D radius apertures).
\item[\textbf{Model 4}]: RF trained using the combined properties from both Table \ref{Table:1} and Table \ref{Table:2}.
\item[\textbf{Model 5}]: RF trained using the combined properties from both Table \ref{Table:2} and Table \ref{Table:3}.
\item[\textbf{Model 6}]: RF trained using all the properties from Tables \ref{Table:1}, \ref{Table:2}, and \ref{Table:3}.
\end{description}

In the following subsection, we present the results of the RF regression analysis for all six trained RF models listed above.

\subsection{RF Model Results}\label{subsection:3.2}
\begin{figure}[h]
    \centering
    \includegraphics[scale=0.4]{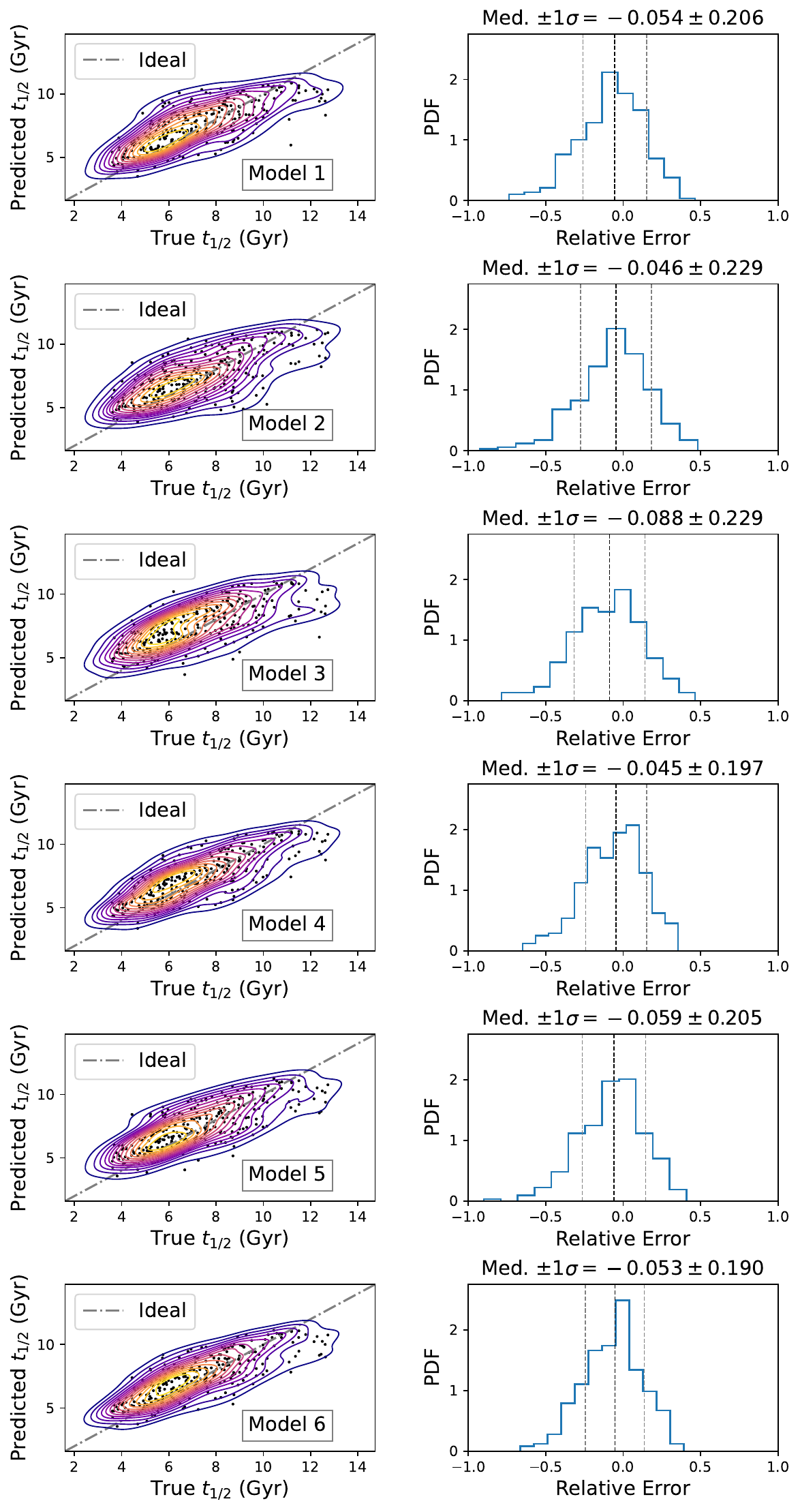}
    \caption{The left panel illustrates the relationship between the true and the predicted $t_{1/2}$ values 
    using the six trained RF models through a bi-variate joint distribution or 2D probability density function. The black data points represent the scatter between the RF predictions and the true $t_{1/2}$ values for the test dataset relative to the ideal case of perfect predictions, depicted by a 45-degree dotted grey diagonal line. The right panel shows the PDF for the relative errors between the true $t_{1/2}$ and the predicted values by the RF models, along with the median error and the standard deviation, $1 \sigma$.}
    \label{fig:3}
\end{figure}

In this subsection, we present the results of the RF algorithm.  We assess the accuracy of the $t_{1/2}$ predictions against the true $t_{1/2}$ for the six RF models described above. The results are shown in the left panels of Figure \ref{fig:3}. These panels provide a two-dimensional joint density distribution of the true versus RF-predicted $t_{1/2}$ values, allowing us to investigate the scatter in our predictions relative to the ideal one-to-one model, depicted by a dotted grey line. 

Additionally, we quantify the relative error between the RF-predicted $t_{1/2}$ values and the actual $t_{1/2}$ values of the test dataset, which is shown on the right panels of Figure \ref{fig:3}. The relative error $t_{1/2}$ values denoted by $t_{1/2}$ within the plots is defined as \(\left(t_{1/2,True} - t_{1/2,Predicted}\right) / t_{1/2,True}\). We first observed that the joint density distribution between the predicted and actual $t_{1/2}$ values 
tends to have a different slope 
relative to the diagonal line and that the average predicted values are slightly higher than the true ones.
This is also evident from the median relative error values, which have a small negative offset from zero in the relative error histogram plot (Figure \ref{fig:3}, right panel). This offset from the ideal model is largest for Model 3 (linked to the properties of Table \ref{Table:3}) and smallest for Model 2 and Model 4. Thus, by assessing all six models, we infer that they overpredict the actual $t_{1/2}$ values of the test sample, with a median bias ranging from 4\% to 9\%.

To further assess the robustness of the RF predictions for our six models, we analyze the joint distribution relative to the diagonal line and the standard deviation of the relative error histogram. A tight clustering of the joint distribution along the diagonal line would indicate high accuracy and low prediction errors.
Even though all the models closely follow the diagonal line, the remaining spread suggests that incorporating additional features could further reduce this spread and enhance the accuracy of predictions.
The relative error histograms for the six models exhibit variability in the predictions. The standard deviation estimates for the six models indicate that errors in the RF predictions can deviate from the median relative error in the range of 19\% to 23\%. From Figure \ref{fig:3}, we observe that the width of the spread of the joint distribution and the standard deviation of the relative error histogram is lowest for Model 6, which uses all the properties from Tables \ref{Table:1}, \ref{Table:2}, and \ref{Table:3} for training.

As mentioned in the previous subsection, we incorporated sample weights while training the models to address the under-sampling issue at the very low and high ranges of $t_{1/2}$ values. However, we still observed over-prediction for very early-forming halos and under-prediction for very late-forming halos across all six cases considered in our analysis, attributed to undersampling at those extremes.

During the training of the Random Forest (RF) algorithm, each decision tree is trained using a bootstrapped sample of the training dataset. On average, approximately 63.2\% of the original training data is included in the bootstrapped sample for training each tree, while the remaining 36.8\%, referred to as out-of-bag (OOB) data, is excluded from the training of that specific tree.
For each data point in the OOB sample, a corresponding prediction exists within the \texttt{OOB\_prediction\_} array from scikit-learn. This prediction is obtained by aggregating and averaging the predictions from the decision trees within the RF that did not use this data point for training. The OOB prediction data can be used to evaluate the performance metrics of the trained models without the need for a separate validation set.

\begin{figure}[h]
    \centering
    \includegraphics[scale=0.42]{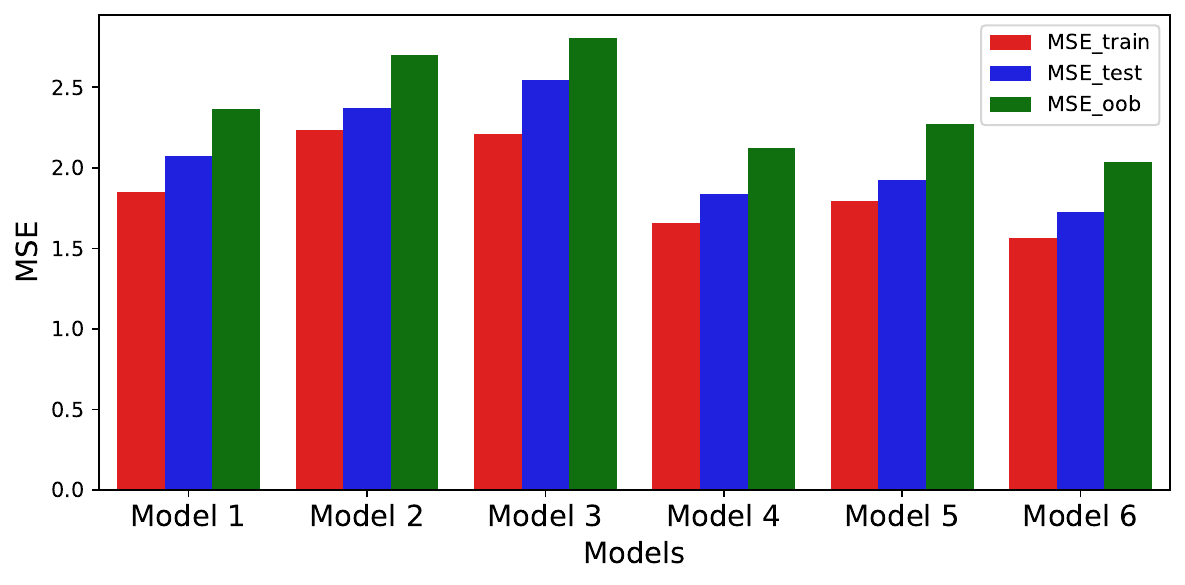}
    \caption{Mean squared error (MSE) for the training, test, and out-of-bag samples for the six models trained using the RF algorithm to assess the model performance.}  
    \label{fig:4}
\end{figure}

To further compare the performance of the six RF models, we evaluated and plotted the mean squared error (MSE) for the training dataset, testing dataset, and OOB sample using the predictions from each trained RF model in Figure \ref{fig:4}. The plot shows that the mean squared error for the training set for all six models is lower than the test and OOB sample mean squared error, indicating effective training. However, the mean squared error for the training dataset is not drastically lower than the test and OOB sample mean squared error, suggesting minimal overfitting. Overall, for all six models, the mean squared error for the test dataset closely aligns with the mean squared error for the training and OOB datasets, indicating good generalization. 
Among all the models, Figure \ref{fig:4} illustrates that Model 6, which utilizes all the data from Tables \ref{Table:1}, \ref{Table:2}, and \ref{Table:3}, exhibits a significant reduction in training, testing, and OOB sample mean squared error compared to the other trained models. The OOB MSE is slightly higher than the test dataset MSE due to sampling differences in both methods. The test set predictions use all trees in the forest which leads to a more stable and typically lower MSE. On the other hand, OOB predictions only use the subset of trees that did not see the particular data during training, resulting in a slightly higher MSE due to the smaller and potentially less representative OOB sample.
 
\begin{figure}[h]
    \centering
    \includegraphics[scale=0.42]{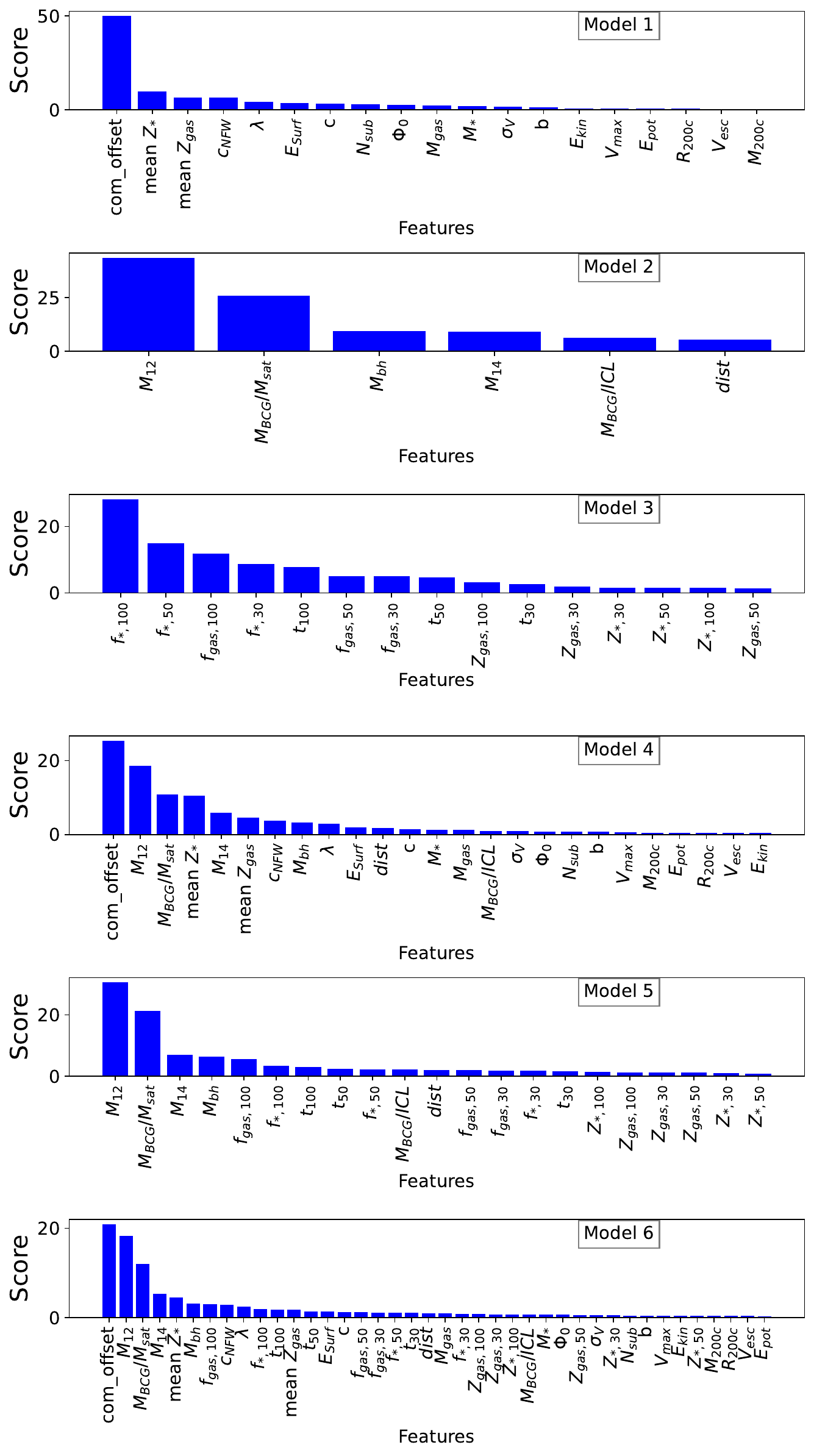}
    \caption{Permutation importance of the features in the six different random forest regressor models we considered in this study.}
    \label{fig:5}
\end{figure}

Lastly, the importance of each feature for the six models we trained in our study was determined using the \texttt{scikit-learn} \texttt{permutation\_importance} function \cite{breiman2001random}. The algorithm works by first calculating the baseline performance of the model using a metric (such as mean squared error) on the dataset. We then select one feature at a time and permute its values to break any relationship between the feature and the target values. The model is then evaluated with the shuffled feature, and its performance is compared to the baseline model performance to gauge the importance of the feature. This process is applied recursively to all the features in the dataset. In Figure \ref{fig:5}, we display the permutation feature importance of the six models considered in this study. 
The first thing we notice from Figure \ref{fig:5}  
is that the most important feature is overall the \textit{com\_offset} of the halo, for which we found a strong positive correlation with the $t_{1/2}$, with a Pearson's correlation coefficient value of 0.63. The \textit{com\_offset} is an indicator of the dynamical state of the halo \citep{ludlow2012dynamical,mann2012x} and, it's known that relaxed halos tend to assemble their mass earlier compared to unrelaxed ones \citep[see, e.g.,][for more discussion on the correlation between a halo's dynamical state and its formation time]{mostoghiu2019three, power2012dynamical}.  
Following, the other important properties contributing to the statistical performance of Model 6 are $M_{12}$, $M_{BCG}/M_{sat}$, $M_{14}$, $\textit{mean} Z_{*}$, $M_{bh}$, and so on. We observe a strong negative correlation between the magnitude gaps ($M_{1,2}$, $M_{1,4}$) and $t_{1/2}$, with Pearson's correlation coefficients of $-0.63$ for $M_{12}$ and $-0.56$ for $M_{14}$, respectively.  
As discussed in the introduction, that trend was expected since the hierarchical growth of the BCG can induce a change in the magnitude gap $M_{12}$ or $M_{14}$ which could serve as an indicator of $t_{1/2}$ \citep{golden2018impact,golden2019impact,golden2022observed,golden2024hierarchical}.
We also note here that the mass gap (or magnitude gap in observations) is also strongly anti-correlated with the \textit{com\_offset}, and has been used as a parameter for quantifying the halo dynamical state.

We have so far utilized halo properties extracted from AHF (Table \ref{Table:1}), properties corresponding to the BCG-ICL region (Table \ref{Table:2}), and stellar and gas properties of the BCG at different aperture cuts (Table \ref{Table:3}) to fit the RF models for predicting $t_{1/2}$. The analysis of the six random forest models in this section shows that the random forest algorithm can indeed predict the formation time of the halos and the model's prediction accuracy extensively improves as we incorporate more features into the model
However, the feature importance plot in Figure \ref{fig:5} also indicates that not all properties play an important role in predicting $t_{1/2}$. Therefore, instead of naively expanding our feature space by including all possible properties, in the next section, we will focus on using the BCG properties that can be computed from observations. This approach will reduce the complexity of our RF model by narrowing down our feature space.

Moreover, when defining the BCG region, we discussed how the radial aperture size used to define the BCG is selected arbitrarily in different literature, and in this study, we utilized three different aperture radii to compute the BCG's stellar and gas properties, all of which are detailed in Table \ref{Table:3}.
However, if the chosen aperture radius is too small, we may miss the region that is more important for predicting $t_{1/2}$. On the other hand, though $f_{*, 100}$ is the most important feature in Model 3 as expected, choosing a larger aperture might introduce a lot of unimportant information, which could overshadow the region most helpful in predicting $t_{1/2}$.
To derive the most important and clean information for $t_{1/2}$ around the BCG as suggested by \cite{golden2022observed, golden2024hierarchical}, in the next section, we will utilize different CNN models that rely solely on the stellar and gas properties of the BCG region for each halo in our dataset.
After training these CNN models, we will use them to identify robust radial aperture ranges for each BCG baryonic property that most strongly correlates with the formation time of the halos.
Using the identified aperture ranges for the baryonic properties from the CNN models' results, we can directly compute these properties and train random forest models with features selected based on prior informed decisions rather than using them arbitrarily.

\section{Convolutional Neural Networks}\label{section:4}
Convolutional Neural Networks (CNNs; \citet{lecun1998gradient, lecun2015deep}, for a detailed review see \citet{aloysius2017review}) are a class of Artificial Neural Networks (ANNs) commonly used for image-based machine learning problems, where the typical input to the network involves two-dimensional image data.
In contrast to traditional ANNs, where each neuron in a layer is interconnected to every neuron in the previous layer, CNNs first process the input image through layers dedicated to convolution and pooling operations, which are subsequently followed by fully connected layers of neurons.
A single convolutional layer of a CNN consists of multiple kernels,  each having a lower dimensionality than the input data. As these kernels traverse the image, they perform convolution by calculating the scalar product between the filter elements and the corresponding input elements, thereby generating a feature map. The kernels are responsible for extracting patterns from the input two-dimensional data, such as horizontal and vertical textures. The components within each kernel constitute a set of weights that are fine-tuned during the CNN training process.

The generated feature map is then activated by an element-wise activation function, such as the rectified linear unit (ReLU, \citet{nair2010rectified}), which introduces non-linearity into the CNN, enabling the model to learn complex patterns. Mathematically, RELU is defined as \( f(x) = \text{max}(0,x) \), which ensures that all negative values (representing black pixels) are ignored, preserving only positive values (corresponding to gray and white pixels).

The activated feature map from the convolutional layer is then processed by the pooling layer, which reduces the dimensionality of the feature map while retaining the most important information. This is typically achieved by applying either max pooling or average pooling. In the case of max-pooling, the operation extracts the maximum element from the area of the feature map covered by the pooling filter, whereas, in average pooling, it calculates the average of all elements in the region of the feature map covered by the pooling filter.
As the feature map passes through convolutional pooling layers, it extracts larger features and reduces in size, cutting computational costs while retaining key information. The output is then flattened to a one-dimensional layer for fully connected layers. The final output can be a single neuron for binary classification or regression, or multiple neurons for multiclass classification.

The network consists of several hyperparameters that can be tuned to achieve optimal performance. We only considered three hyperparameters in our study which are as follows:  
\begin{itemize}
    \item \textit{epochs}: It refers to the number of complete passes of the training dataset through the model. One complete pass involves processing the data set forward and backward through the network once, ensuring that the model has seen and learned from every training example exactly once. 
    \item \textit{learning rate}: It specifies the magnitude of the step taken in the parameter space by the optimization algorithm while adjusting the weights of the CNN during the training process.
    \item \textit{dropout rate}: It is a regularization technique that prevents overfitting in the model by ignoring some neurons in the fully connected layer. The dropout rate represents the fraction of neurons that will be randomly ignored during training.
\end{itemize}
In the next subsection, we detail the creation of input maps/images for the training, the CNN architecture, and the optimization hyperparameters.

\subsection{CNN Architecture and Training Setup}\label{subsection:4.1}

\begin{figure}[h]
    \centering
    \includegraphics[scale=0.5]{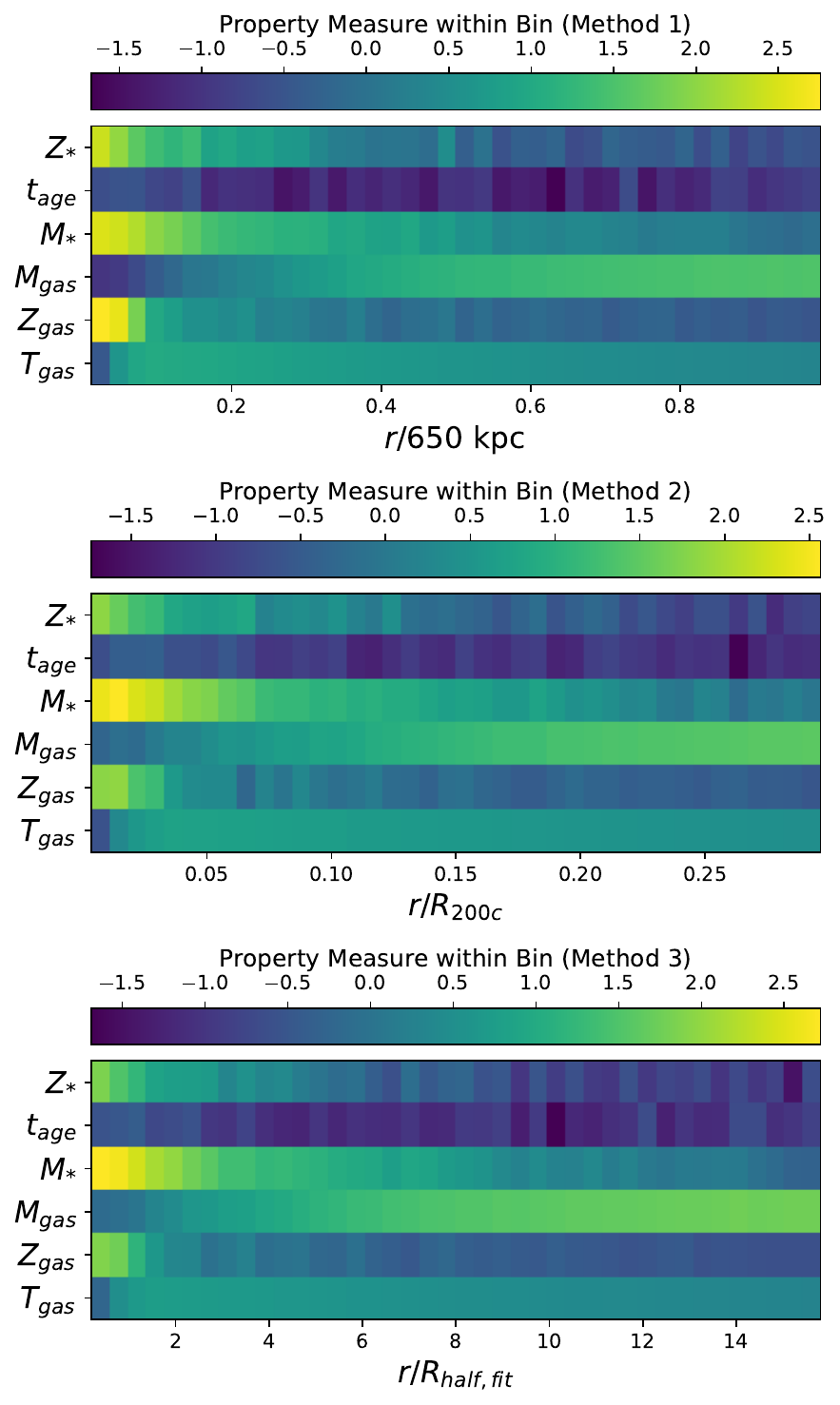}
    \caption{
Property maps for a halo with \(R_{200c} = 1035.84 \, \text{kpc}/h\) and \(M_{200c} = 2.58 \times 10^{14} h^{-1} M_{\odot}\) with the three different binning methods. The \(M_{*, \text{half}}\) and \(R_{\text{half}}\) values for this halo from the CAESAR catalog are \(6.85 \times 10^{11} h^{-1} M_{\odot}\) and \(23.18 \, \text{kpc}/h\), respectively, while the half-mass radius estimated from the linear fit is \(R_{\text{half, fit}} = 26.67 \, \text{kpc}/h\).
The upper panel presents the property map for this halo using binning method 1, i.e., forty uniformly spaced bins from the halo center to a fixed distance of 650 kpc. The middle panel shows the property map for the same halo using binning method 2, i.e., forty uniformly spaced bins from the center of the halo to \(0.3 \times R_{200c}\). The lower panel depicts the property map for the same halo using binning method 3, i.e., forty uniformly spaced bins from the center of the halo to \(16 \times R_{\text{half, fit}}\).
}
    \label{fig:6}
\end{figure}

An initial prerequisite for using a CNN is the input image data.
We first derived property maps for the halos used in the RF analysis, which will be used as input images for training the CNN network. The training and test samples from the RF study were used directly, meaning no new train-test split was created. To produce property maps, we considered the stellar and gas properties of the halos, which were computed based on their respective particle information within the simulation. The properties we considered include: (i) mass-weighted stellar metallicity (Z$_{*}$), (ii) mass-weighted stellar age ($t_{\text{age}}$), (iii) stellar mass (M$_{*}$), (iv) gas mass (M$_{gas}$), (v) mass-weighted gas metallicity (Z$_{gas}$), and (vi) mass-weighted gas temperature (T$_{gas}$), which were computed using the information of the stellar and gas particles within the simulated halos. For each halo, the gas and stellar particles were binned into forty uniformly distributed bins 
in three different ranges of radial distances explained below.
The six properties stated before were calculated within each bin to generate a property map of size $6 \times 40$. Note that only the particles belonging to the halo, and not any subhalos within it, are considered for creating these property maps.
The three different methods used in this study for binning the stellar and gas particles are as follows:
\begin{itemize}
    \item \textbf{Binning Method 1}: 40 uniformly spaced bins defined from the center to $650$ kpc physical distance to ensure a consistent particle distribution.
    
    \item \textbf{Binning Method 2}: 40 uniformly spaced bins defined from the halo center to a radius of \(0.3 \times \text{R}_{200c}\).
    
    \item \textbf{Binning Method 3}: 40 uniformly spaced bins defined from the center of the halo to \(16 \times \text{R}_{\text{half, fit}}\), where \(\text{R}_{\text{half, fit}}\) is the half-stellar mass radius of the central galaxy within the halo. The \(\text{R}_{\text{half, fit}}\) values for our halo dataset are based on the CAESAR catalog. However, the CAESAR catalog's radius estimations showed a significant scatter for a fixed value of stellar mass within the half-stellar mass radius (i.e. $M_{*, \text{half}}$), with a deviation of approximately 8.78 kpc from the best linear fit, which results in the ML prediction with very large scatter. Therefore, we used the half-mass radius values predicted by the linear fit relation between the $M_{*, \text{half}}$ and \(\text{R}_{\text{half}}\) of the CAESAR catalog.
\end{itemize}
The bin positions within the map for each halo in the first binning method are always normalized to the maximum bin edge, i.e., 650 kpcs. In the latter two cases, they are normalized to their corresponding \(R_{200c}\) and \(R_{\text{half, fit}}\), to place all the halos used in the analysis on equal footing.
In Appendix \ref{Appendix:1}, we present the distribution of $0.3 \times R_{\mathrm{200c}}$ and $16 \times R_{\mathrm{half,fit}}$ for all the halos in Figure \ref{fig:A.1} to give an idea about the average spatial resolution of the Binning Method 1 and Binning Method 2.  
The maps developed for each halo correspond to a 2D image, where each vertical slice of the property map represents a bin and consists of six vertical columns of pixels, with their pixel values corresponding to the stellar and gas properties of the halo within that bin.
For some halos, we noticed that gas properties were computed as \texttt{NAN} due to the absence of gas particles within the bins. We excluded those bins from the property maps so that it does not lead to CNN training failure.
We could not ignore the halos with this issue because a few bins (1 or 2) with \texttt{NaN} values 
occur frequently in some halos within our dataset. If we ignored these halos, our sample size would become very small, making it unsuitable for training the CNN network. However, they are resized later on to have the same dimensions to make CNN training possible.
Also, the numerical value ranges for the six physical properties in each row can vary widely. This leads to pixel value distributions that differ significantly between properties. In our case, the gas temperature, stellar mass, and gas mass values are considerably larger than the other three properties, causing these three properties to overshadow the rest. We applied transformations to these maps to ensure that the magnitude range difference between the six properties does not impact the CNN training and that each property's overall distribution across all maps in the combined training and test sets is close enough to a normal distribution. A base-10 logarithmic transformation was applied to compress the large magnitudes of the gas temperature, stellar mass, and gas mass within the property maps. The gas metallicity values within the bins across all the maps in the training and testing data set are between 0 and 0.075 and the distribution for it is very squeezed for smaller values range. Therefore, to expand the distribution and to make it more distinct we performed a square root transformation over all the values of gas metallicity. The stellar age values were transformed using:
\begin{align}
t_{\text{age}} = \log_{e} \left( \text{max}(t_{\text{age}}) + 1 - t_{\text{age}} \right)
\end{align}
where \(\text{max}(t_{\text{age}})\) represents the maximum stellar age across all the maps, i.e., after concatenating the whole training and testing map datasets. This transformation ensures that the overall distribution of stellar age values is not skewed toward the tail end and approximately follows a normal distribution. No transformation was applied to the stellar metallicity data within the map. Finally, after applying the transformations to the individual physical properties of the maps, we applied Z-score normalization to each property. The Z-score normalization is mathematically expressed as:
\begin{align}
X_{p}^{'} = \frac{X_{p} - \mu_{p}}{\sigma_{p}}     
\end{align}
where \(p\) corresponds to one of the six physical properties used in the maps. $X_{p}$ represents the original value of property \(p\), while $X_{p}^{'}$ is the normalized value. $\mu_{p}$ and $\sigma_{p}$ are the mean and standard deviation of property $p$, calculated across the entire dataset by combining all the maps from the training and test sets. The Z-score normalization scales the data with a mean of zero and a standard deviation of one, ensuring that each feature or property is placed on equal footing, regardless of the scale of the original data.
All the maps are then resized using OpenCV \citep{opencv_library} to have the dimensionality of $6 \times 40$ so that the input map sizes for CNN training are consistent.
Like the halo dataset from the random forest analysis, we now possess 1,630 property maps in the training dataset and 288 in the testing dataset.
In Figure \ref{fig:6}, we present the property map for a halo using the three different binning methods, with all normalization, and Z-score standardization applied to each row or property within the map.

\begin{figure*}[h]
    \centering
    \includegraphics[scale=0.45]{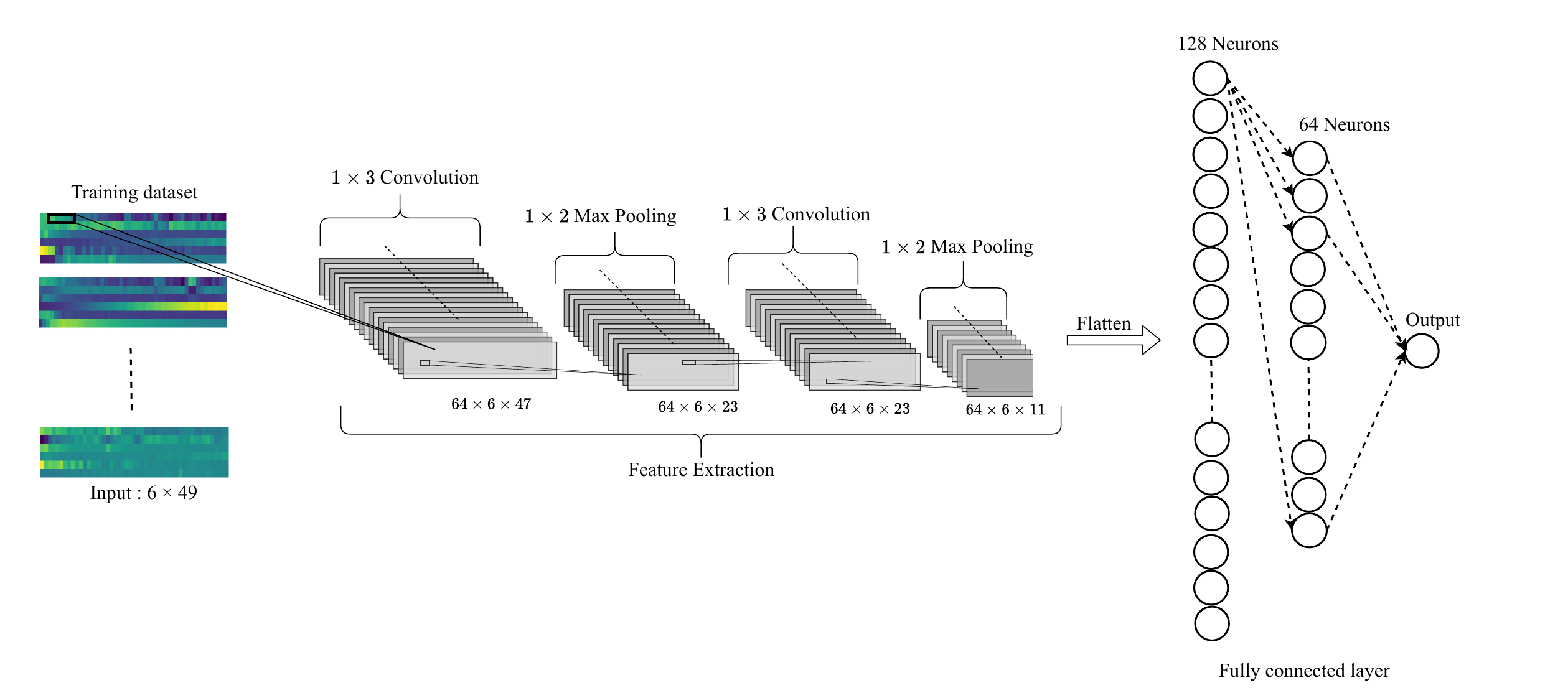}
    \caption{Architecture of the `single-channel' CNN used in our study. Our network employs two convolutional pooling layers for feature extraction and two fully connected layers for parameter estimation.}
    \label{fig:7}
\end{figure*}

After computing these maps for the halo dataset, we designed our network to use these property maps to make $t_{1/2}$ predictions.
We implemented our CNN architecture using the Keras package with Tensorflow backend, written in Python. The architecture of our ‘single-channel’ CNN is as follows:
\begin{itemize}
    \item \textbf{Convolutional-Pooling Layer 1}
    \begin{enumerate}
        \item $1 \times 3$ convolution, $64$ filters, with ReLU activation
        \item $1 \times 2$ stride-2 max pooling
    \end{enumerate}
    \item \textbf{Convolutional-Pooling Layer 2}
    \begin{enumerate}
        \item $1 \times 3$ convolution, $64$ filters, with ReLU activation. The padding is set to `same' so that the output feature map has the same spatial dimensions as the previous layer feature map. 
        \item $1 \times 2$ stride-2 max pooling
    \end{enumerate}
    \item \textbf{Flattening Layer}
    \item \textbf{Dense Layer 1}
    \begin{enumerate}
        \item 128 neurons, fully connected followed by activation function set to ReLU
    \end{enumerate}
    \item \textbf{Dense Layer 2}
    \begin{enumerate}
        \item 64 neurons, fully connected followed by activation function set to ReLU
    \end{enumerate}
    \item \textbf{Output Layer}
    The final output layer uses the linear activation function.
\end{itemize}
We chose 1D vector kernels for convolution and pooling operations rather than 2D kernels because different rows in the input maps correspond to different physical quantities which are independent of each other, and we do not want them to convolve with each other by using 2D kernels either.  
In Figure \ref{fig:7}, we give the pictorial representation of our `single channel' CNN network.
Instead of using different CNN architectures for each of the three binning methods, we will use the same architecture to see how the predictions are affected by binning gas and stellar particles using different criteria. We trained three CNN models which are as follows:
\begin{description}
\item[\textbf{CNN Model 1}]: CNN trained using input training maps generated with Binning Method 1.
\item[\textbf{CNN Model 2}]: CNN trained using input training maps generated with Binning Method 2.
\item[\textbf{CNN Model 3}]: CNN trained using input training maps generated with Binning Method 3.
\end{description}

The range of values for the hyperparameters we consider are as follows:
\begin{itemize}
    \item \textit{epochs} : $[17, 21, 31, 41, 49]$
    \item \textit{dropout rate} : $[0.1, 0.2, 0.3, 0.4, 0.5]$
    \item \textit{learning rate} : $[ 10^{-5},10^{-4},10^{-3},10^{-2},10^{-1}]$
\end{itemize}

To tune the hyperparameters of these CNN models, we used \texttt{GridSearchCV} again with five-fold cross-validation.
We used the Adam Optimiser \citep{kingma2014adam} with a mean squared error loss function during the training of our CNN model. The training data were divided into batches of 51 training maps each.
The weighting criteria based on the distribution of $t_{1/2}$ in Figure \ref{fig:2} are also applied again during the training of the CNN models.

\subsection{CNN Results}\label{subsection:4.2}
\begin{figure}[h]
    \centering
    \includegraphics[scale=0.42]{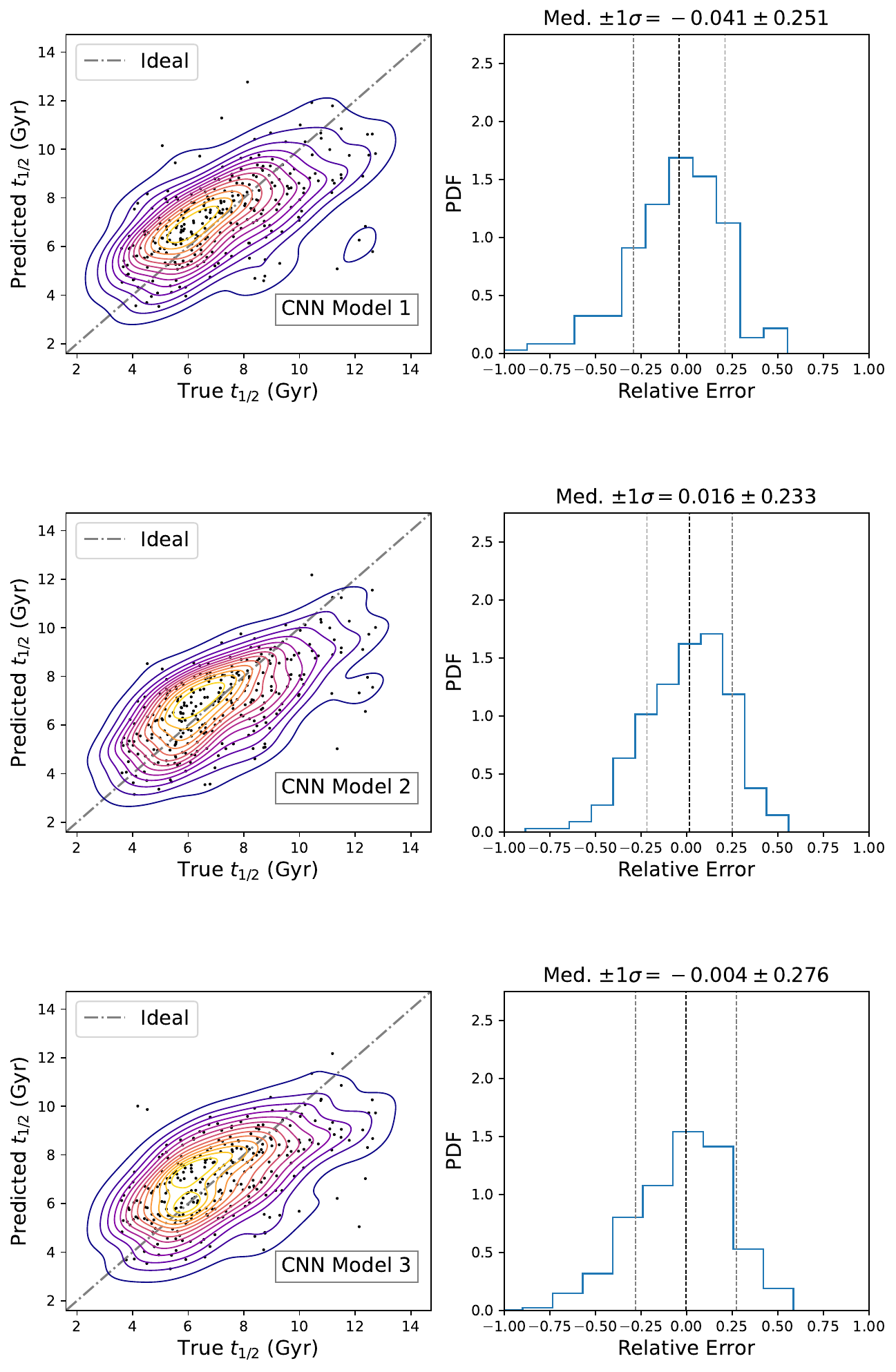}
    \caption{
The $t_{1/2}$ predictions and their accuracy assessment for all three CNN models. The first panel corresponds to the model that used binning Methodology 1, the middle panel to binning Methodology 2, and the last panel to binning Methodology 3. The left column of the plot shows scatter plots (i.e. depicted by black points in the scatter plots) of the true $t_{1/2}$ values for the test dataset versus those predicted by the CNN models. The contour plots depict the joint 2D distribution between the true and predicted $t_{1/2}$ values, with the 45-degree diagonal line representing the ideal model prediction. The right panel shows the PDF of the relative errors between the true $t_{1/2}$ values and those predicted by the CNN models, highlighting the median error and the standard deviation, $1 \sigma$.
    }
    \label{fig:8}
\end{figure}

Our predictions for halo formation times in the test set for the three different binning methodologies are shown in Figure \ref{fig:8}. The left column of the figure presents scatter plots comparing the CNN-predicted $t_{1/2}$ to the true $t_{1/2}$ for each of the three binning criteria. The contour plot illustrates the 2D joint density distribution of the CNN-predicted and true $t_{1/2}$ values for the test set. The right column contains histogram plots showing the relative error between the CNN predictions and the true $t_{1/2}$ values, again for the three different binning methodologies used to create the maps for the halo dataset. 

We infer from the relative error histograms for CNN Model 1 and CNN Model 3 that the model overestimates the formation time values, with median relative error biases of 4.1 \% and 0.4 \%, respectively. This is low compared to all the random forest models discussed in Section \ref{section:3}. On the other hand, the predictions for the CNN Model 2 underestimate the formation time values, with a median relative error bias of 1.6 \%. We see that the median relative error bias is the lowest for the CNN model 3 where the input maps for training and testing are generated following binning Methodology 3, i.e., 40 uniform bins located between 0 to \(16 \times R_{\mathrm{half,fit}}\).

However, the scatter of the relative errors for the CNN predictions is slightly larger compared to all the random forest models investigated in subsection \ref{subsection:3.2}. The same can be inferred from the width of the 2D joint distribution in the left panel plots for all three methods in Figure \ref{fig:8}.
The scatter in the relative errors of $t_{1/2}$ predictions from median relative errors is the lowest for CNN Model 2, where the maps are generated by binning the gas and stellar particles within 40 uniformly spaced bins from 0 to $0.3 \times R_{200c}$.
Another contrast in the CNN models' relative error histogram compared to the RF models' is that the left tail of the CNN models' histogram is skewed, implying that the CNN models tend to overpredict the formation time values by relatively large errors.
The promising aspect, however, is that with just six observable properties around the central galaxy hosted within the dark matter halo—thereby reducing our feature space—we can achieve predictions with a reduced median relative error bias, though with a scatter that is only 25\% larger compared to the random forest models.

Additionally, we interpreted the CNN models using saliency maps and analyzed them to identify the radial range that exhibits the strongest correlation with galaxy properties, enabling its direct incorporation into a machine-learning model for estimating the halo formation time. However, we did not include the saliency maps analysis in the main section of the manuscript, as the saliency maps are not easily interpretable due to the high correlation between features in different bins and their generally low correlation with 
\(t_{1/2}\) itself at different radii. An extensive discussion of the saliency maps for the three CNN models trained in this section is provided in Appendix \ref{Appendix:saliency}.

\section{Benchmarking the RF and CNN models}\label{section:correlation and EPS}
In this section, we assess the performance of the RF and CNN models in reproducing the key correlation trends between \(t_{1/2}\) and both \(M_{200c}\) and \(c_{\mathrm{NFW}}\), as well as evaluate the accuracy of their predictions against halo formation time estimates derived from analytical approaches, such as that of \citet{10.1093/mnras/stv689}.

\begin{figure*}[h]
    \centering

    \includegraphics[scale=0.32]{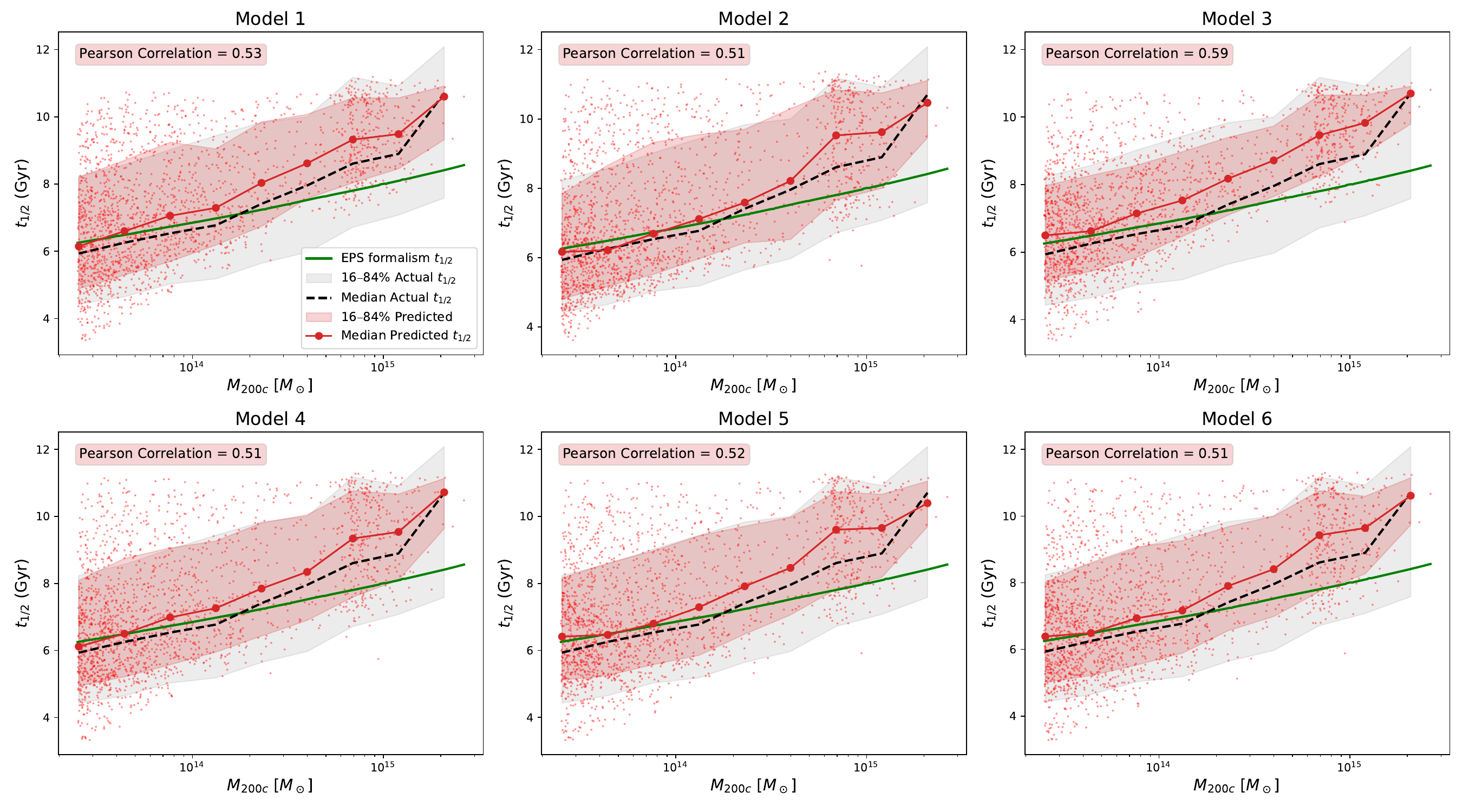}  

    \includegraphics[scale=0.32]{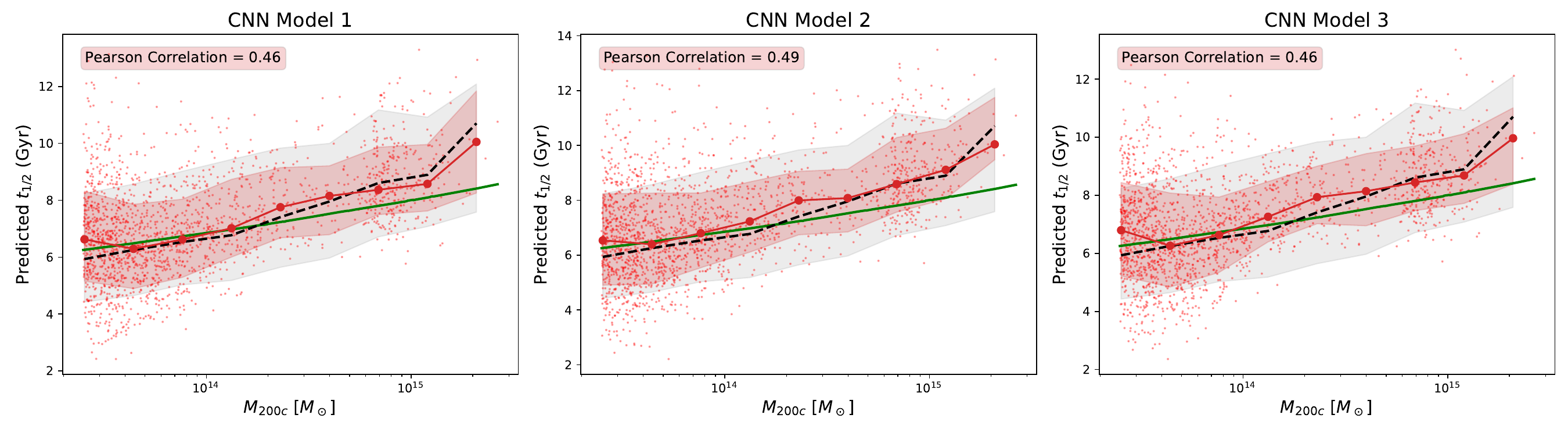}  
 \caption{
Correlation between the halo mass \( M_{200c} \) and \( t_{1/2} \) across different RF and CNN models. The first and second rows show the correlation trends for the six RF models discussed in Section \ref{section:3}, while the third row corresponds to the three CNN models described in Section \ref{section:4}.  
In each panel, the shaded red region represents the 16th–84th percentile range of the ML predictions, and the shaded gray region shows the same range for the actual simulation values. The solid red and dotted black lines indicate the median \(t_{1/2}\) in logarithmic mass bins for the ML models and the simulation, respectively. Pearson correlation coefficients are annotated to quantify the strength of the relationship between \( M_{200c} \) and the predicted formation times. The solid green line gives the \(t_{1/2}\) - \(M_{200c}\) relation based on formation time values calculated using the analytical formalism based on EPS theory as given in \citet{10.1093/mnras/stv689}. 
} \label{fig:correlation M200c}
\end{figure*}

We begin by examining the correlation between the predicted \( t_{1/2} \) and the halo mass \( M_{200c} \). The true relation has already been discussed in Figure \ref{fig:1}. As shown in Figure~\ref{fig:correlation M200c}, all six RF and three CNN models successfully reproduce the expected positive trend between \( M_{200c} \) and \( t_{1/2} \). The median binned relations from the ML-predicted \( t_{1/2} \) values, denoted by red lines in all panels of Figure~\ref{fig:correlation M200c}, align closely with those derived from the true simulation values, shown with dotted black lines. Moreover, the associated \(\pm34\) percentile scatter range shows significant overlap, as the red shaded regions for the ML predictions coincide with the gray regions representing the simulation percentile range as shown in Figure \ref{fig:correlation M200c}. This consistency suggests that the models are effectively capturing the underlying physical correlation. Quantitatively, the overall Pearson correlation coefficient between \( M_{200c} \) and the predicted \( t_{1/2} \) is 0.39, with the RF models achieving a mean correlation of \(\sim 0.52\), and the CNN models \(\sim 0.46\).

\begin{figure}
    \centering
    \includegraphics[scale=0.4]{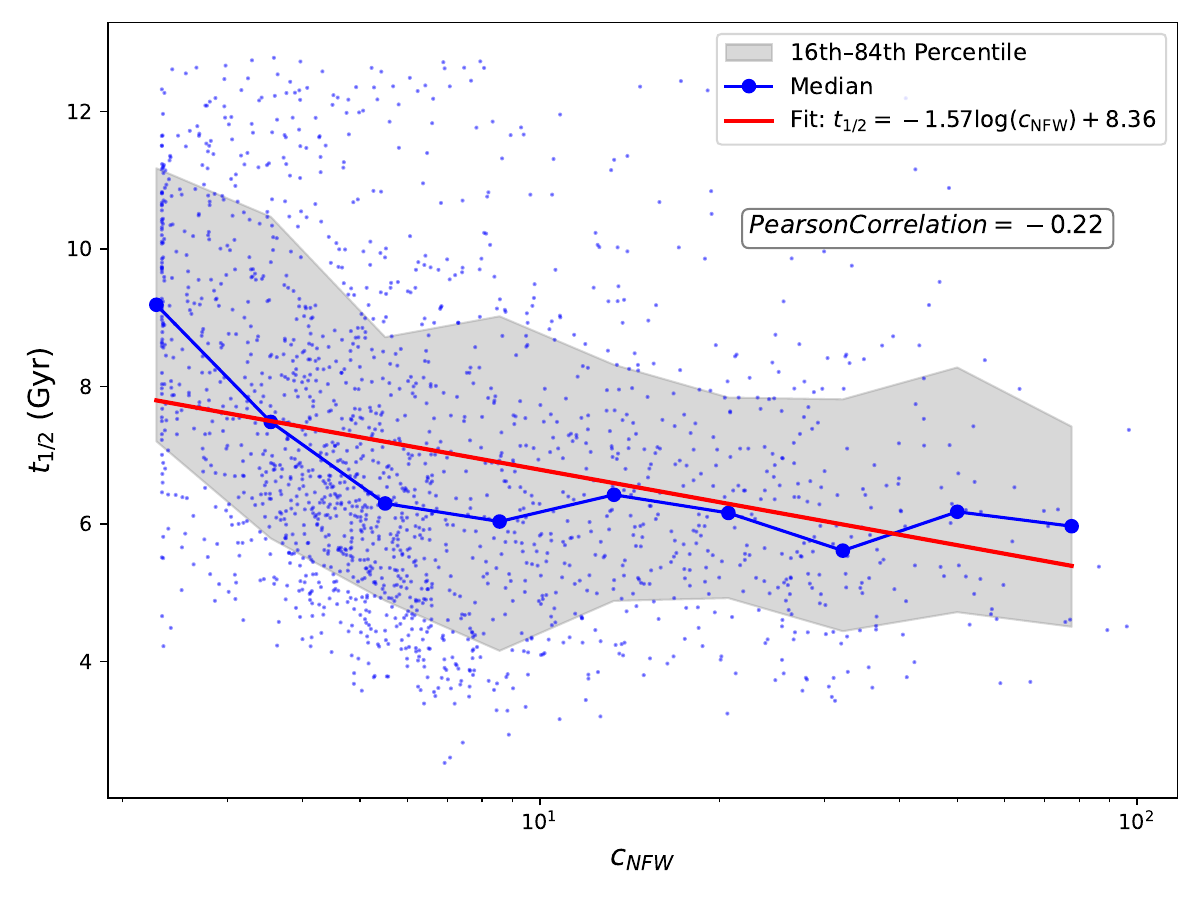}  
    \caption{Scatter plot depicting the relation between the concentration parameter \(c_{\mathrm{NFW}}\) and \(t_{1/2}\) in Gyr for selected sample of halos from The300. The red line depicts the best-fit line obtained by fitting a simple linear relationship between \(c_{\mathrm{NFW}}\) and \(t_{1/2}\).}
    \label{fig:cNFW}
\end{figure}
\begin{figure*}[h]
    \centering
    \includegraphics[scale=0.32]{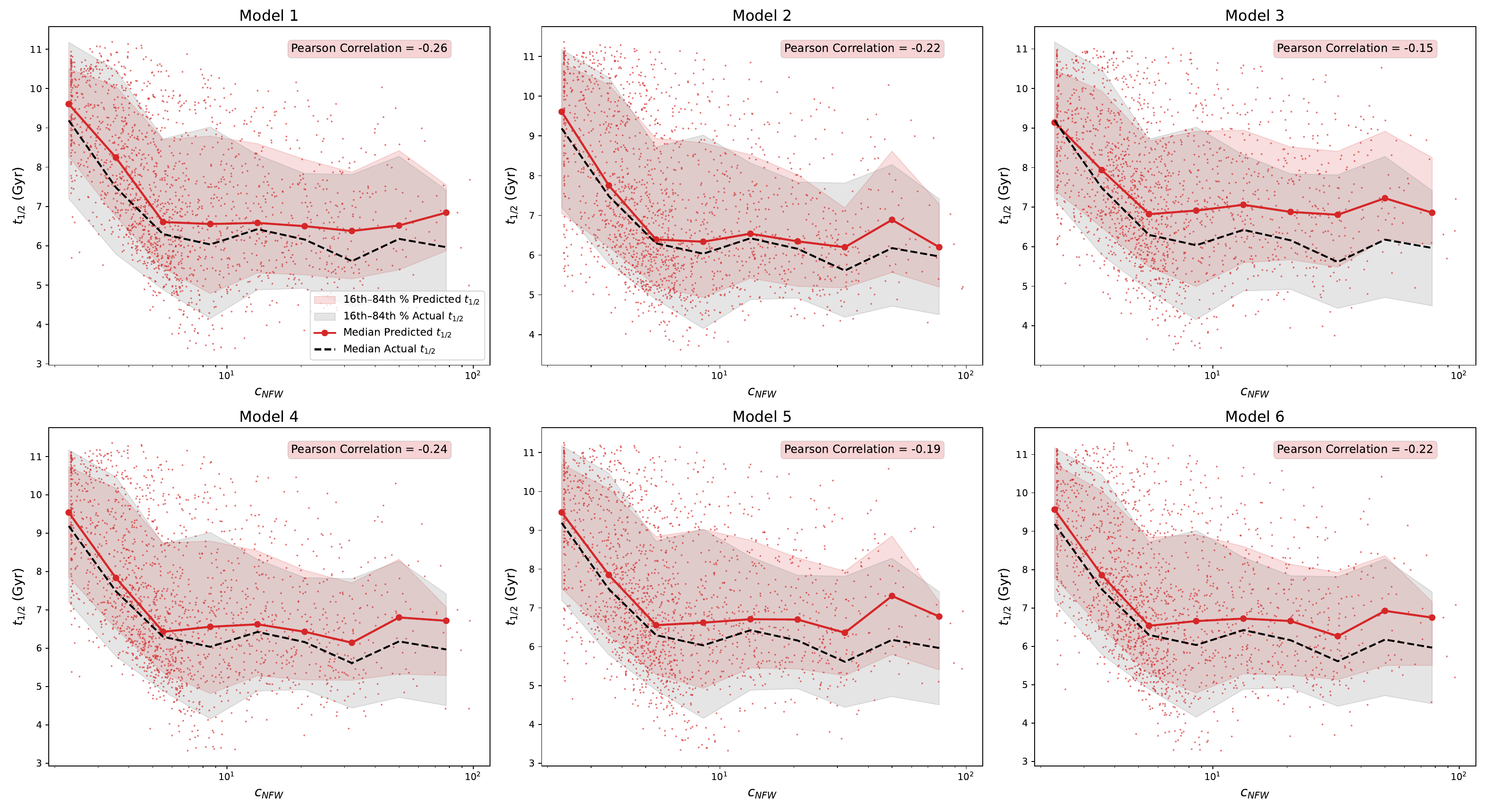} 

    \includegraphics[scale=0.32]{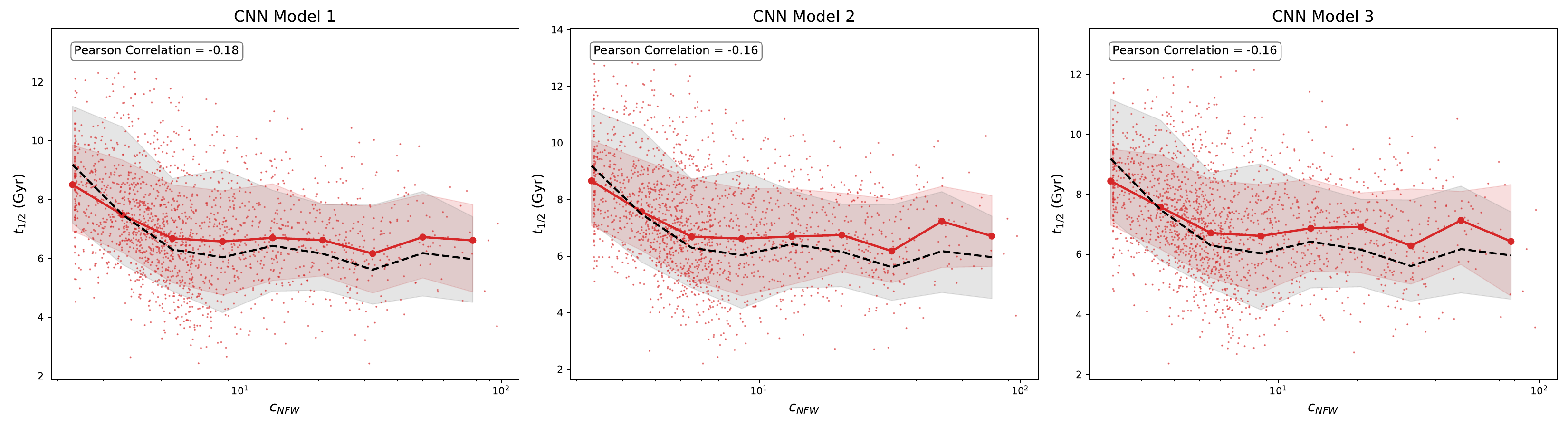} 
 \caption{
Correlation between the halo concentration parameter \(c_{\mathrm{NFW}}\) and \( t_{1/2} \) across different RF and CNN models. The first and second rows show the correlation trends for the six RF models discussed in Section \ref{section:3}, while the third row corresponds to the three CNN models described in Section \ref{section:4}.  
In each panel, the shaded red region represents the 16th–84th percentile range of the ML predictions, and the shaded gray region shows the same range for the actual simulation values. The solid red and dotted black lines indicate the median \(t_{1/2}\) in logarithmic bins for the ML models and the simulation, respectively. Pearson correlation coefficients are annotated to quantify the strength of the relationship between \(c_{\mathrm{NFW}}\) and the predicted formation times.
} \label{fig:correlation cNFW}
\end{figure*}

Turning to the halo concentration parameter \( c_{\mathrm{NFW}} \), we find a clear negative correlation with the true \( t_{1/2} \) of the simulated halo dataset, as shown in Figure \ref{fig:cNFW}. This is consistent with the expectations from hierarchical structure formation, where earlier-forming halos tend to be more concentrated than their late-forming counterparts. As illustrated in Figure \ref{fig:correlation cNFW}, both the RF and CNN models successfully reproduce the expected negative correlation between \( t_{1/2} \) and \( c_{\mathrm{NFW}} \). The true simulation-based correlation is \(-0.22\), while the RF and CNN models yield mean correlations of approximately \(-0.20\) and \(-0.17\), respectively. Similar to the case of \( M_{200c} \), the predicted median trends closely follow the actual median trend, and the percentile ranges largely overlap with those derived from the true values. This indicates that the machine learning models are capable of recovering this inverse relationship.

We also benchmarked the performance of our ML models by comparing the predicted \(t_{1/2}\) values with those obtained from an analytical approach based on the extended Press–Schechter (EPS) theory, specifically following the formulation presented in \citet{10.1093/mnras/stv689}. This assessment was conducted using the \(M_{200c}\)–\(t_{1/2}\) relation shown in Figure \ref{fig:correlation M200c}.
The 16th–84th percentile gray shaded region in Figure \ref{fig:correlation M200c}, derived from the simulation-based \(t_{1/2}\) values, was obtained by tracing the mass accretion histories of halos through merger trees constructed using AHF. These simulation-based formation times exhibit significant scatter when compared with \(t_{1/2}\) values calculated using the \texttt{commah}\footnote{\url{https://bitbucket.org/astroduff/commah/src/master/}} package. This package given in \citet{10.1093/mnras/stv689} implements the analytical framework for modeling the mass accretion histories of halos, from which \(t_{1/2}\) can be computed. The scatter of the simulation data relative to the green line—representing the analytical formalism—highlights the diversity in halo assembly histories: halos of the same mass can have significantly different formation times. 
Nonetheless, the median trend of the simulation-based \(t_{1/2}\) values broadly follows the analytical predictions for halos of lower masses. However, at the high-mass end, the simulation data predict later formation times than the analytical model.
The RF and CNN model predictions show a similar trend and generally follow the simulation results: at lower halo masses, the predicted \(t_{1/2}\) values align well with the analytical estimates, while at higher masses, they diverge, indicating later formation times. However, the RF predictions deviate more strongly from the analytical model than the simulation data, in contrast to the CNN models. Although the CNN predictions also exhibit divergence at high masses, the extent of this deviation is smaller than that seen in the RF models. Overall, this comparison reinforces that both RF and CNN models effectively learn the physical trends of halo assembly from the simulations.

\section{Linear Models for Halo Formation Time Estimation}\label{section:5}
\begin{figure}[h]
    \centering

    \includegraphics[scale=0.4]{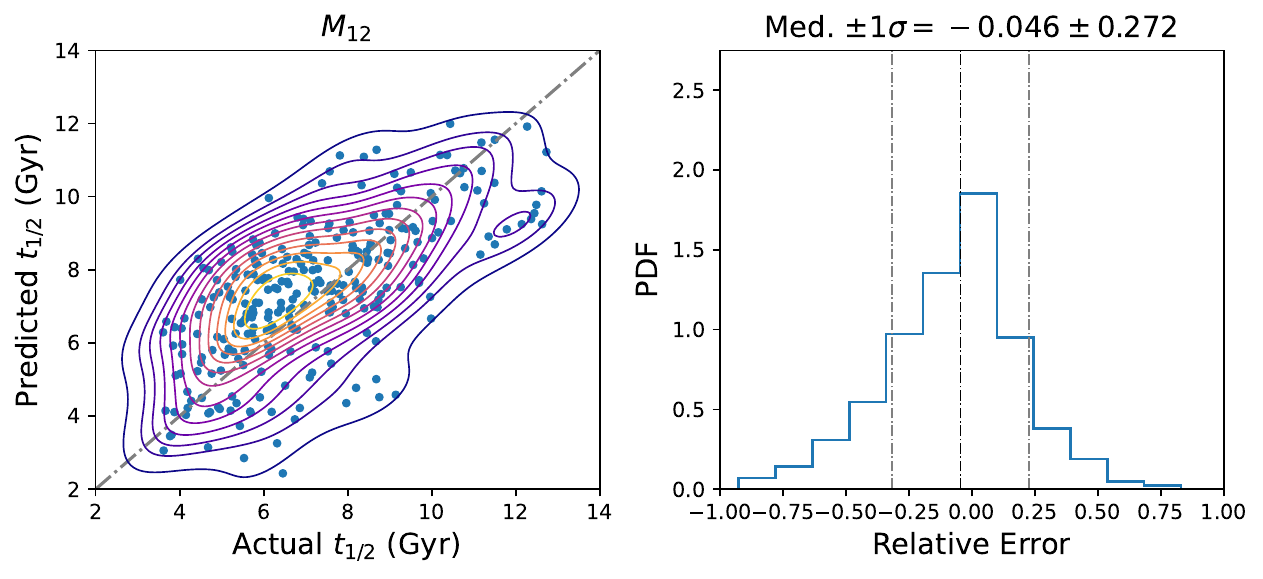}  

    \includegraphics[scale=0.4]{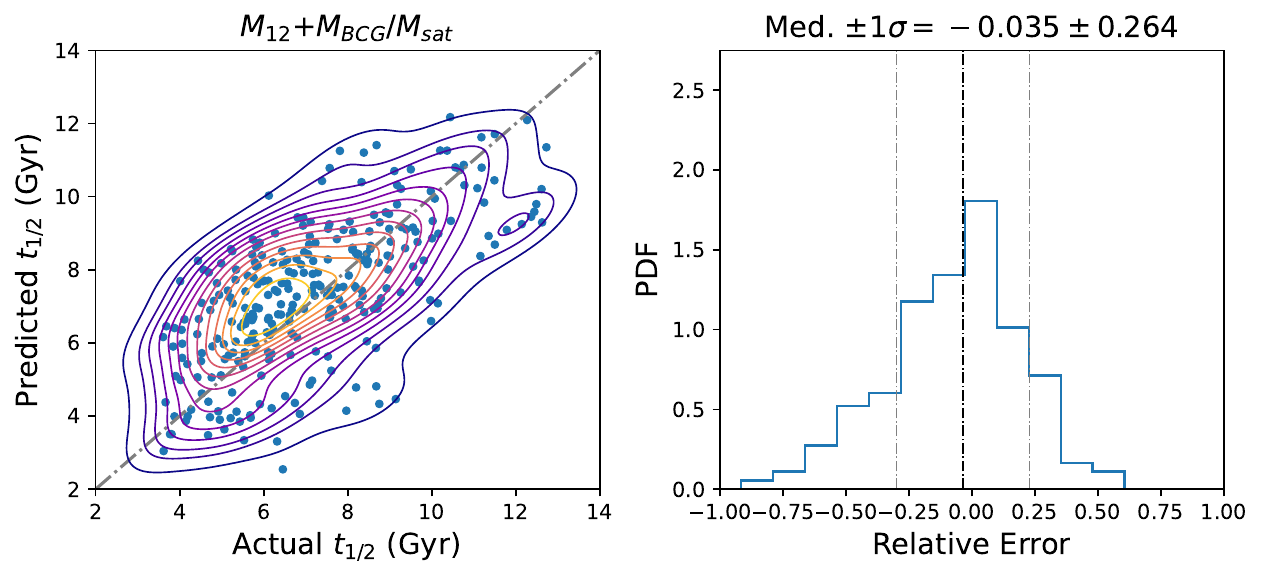}  

    \includegraphics[scale=0.4]{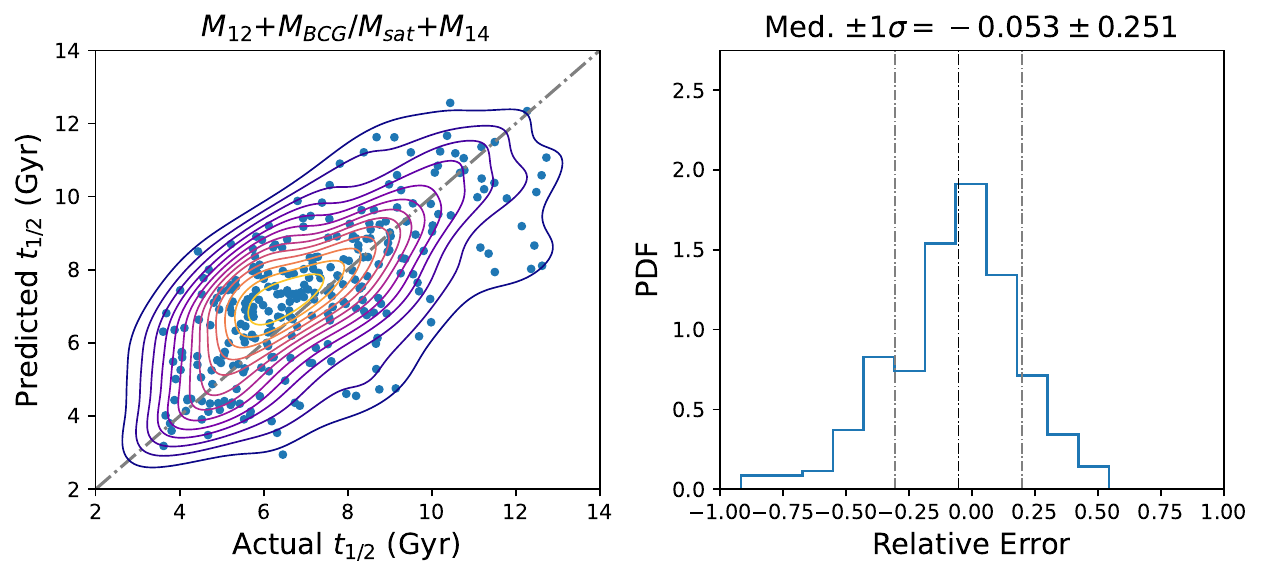}  

    \caption{ 
The first panel compares the \( t_{1/2} \) predictions with the actual values for the test dataset using a linear model based solely on the magnitude gap \( M_{12} \).  
The second and third panels show the linear model predictions when additional observable properties from Tables \ref{Table:2} and \ref{Table:3} in Section \ref{section:2} are incrementally included, namely \( M_{BCG}/M_{sat} \) and \( M_{14} \), along with \( M_{12} \).  
The second panel presents predictions analysis from a linear model using \(M_{BCG}/M_{sat}\) along with \( M_{12} \). Similarly, the third panel show predictions analysis based on \(M_{BCG}/M_{sat}\) and \(M_{14}\)  , respectively, combined with \( M_{12} \).  In the right column, we show the usual scatter and contour density plots comparing the predicted and actual $t_{1/2}$s, while the histogram of relative errors is displayed on the right side.
    }    \label{fig:9}
\end{figure}

In this section, we aim to construct simple linear models that are less complex than CNN and RF models while providing a reasonable estimate of \( t_{1/2} \) without introducing the complexities inherent in advanced machine learning models.  

We begin by building a linear model\footnote{All linear models are based on least squares fitting} to infer \( t_{1/2} \) using only the magnitude/mass gap \( M_{12} \), which represents the logarithmic mass difference between the total BCG mass, defined within a 50 \( h^{-1} \) kpc 3D aperture, and the most massive substructure. As highlighted in the introduction and the feature importance analysis of Model 2, Model 4, and Model 6 of the random forest (discussed in Section \ref{subsection:3.2}), the magnitude gap is a robust indicator of \( t_{1/2} \) \citep{golden2018impact, golden2019impact, golden2022observed, golden2024hierarchical}. Since this quantity can be computed observationally, it provides a practical means of estimating \( t_{1/2} \).  

In the first panel of Figure \ref{fig:9}, we present the \( t_{1/2} \) prediction analysis for the halos in the test set based on a linear model trained using \( M_{12} \) values from the training dataset. The training and test datasets are the same as those used previously.  
This model yields a reasonably accurate estimate of \( t_{1/2} \), with a median relative error that slightly over-predicts values by 4.6 percent and a scatter of 0.271 in the relative error, indicating a moderate spread in predictions. The simple linear relation can be expressed as  
\begin{align}
    t_{1/2} \text{(Gyr)} = -4.26 M_{12} + 8.91 \label{equation:3}
\end{align}  

We further refine this linear model by incrementally incorporating additional properties from the set of observable features listed in Tables \ref{Table:2} and \ref{Table:3} in Section \ref{section:2}.  
The goal is to reduce the scatter and median relative error bias in the \( M_{12} \)–\( t_{1/2} \) relation by integrating more relevant observational properties.  

In the second panel of Figure \ref{fig:9}, we introduce a refined linear model that incorporates \( M_{BCG}/M_{sat} \) alongside \( M_{12} \). The selection of this additional feature is guided by the feature importance plots in Figure \ref{fig:5} and a systematic brute-force approach, where each property from Tables \ref{Table:2} and \ref{Table:3} was tested individually to determine its effect in terms of performance of the \( M_{12} \)–\( t_{1/2} \) relation. Among the tested properties, \( M_{BCG}/M_{sat} \) yielded the best performance in terms of median relative error, bias, and scatter.  

The choice of \( M_{BCG}/M_{sat} \) was based not only on performance improvements but also on its correlation with both \( M_{12} \) and \( t_{1/2} \). While \( M_{BCG}/M_{sat} \) shares some information with \( M_{12} \), it still entails a distinct aspect of the cluster’s properties that influences the linear model’s performance, as seen in Figure \ref{fig:5}. With the inclusion of \( M_{BCG}/M_{sat} \), the scatter in the relative error histogram is reduced to 0.264, and the median overestimation bias decreases to 3.5\%, marking an improvement of approximately 1.1\% over the single-feature model. This refined linear model is expressed as:  
\begin{align}
t_{1/2} \text{(Gyr)} = -4.58 M_{12} + 0.12 {M_{BCG}}/{M_{sat}} + 8.86 \label{equation:4}
\end{align}  
where the definitions of \( M_{12} \) and \( M_{BCG}/M_{sat} \) are provided in Table \ref{Table:2}.  

In the third panel of Figure \ref{fig:9}, we present a further improved linear model that builds upon the previous version by incorporating an additional mass gap, \( M_{14} \), as defined in Table \ref{Table:2}. The inclusion of \( M_{14} \) follows the same brute-force approach, where different observables were tested individually for their impact on the \( M_{12} \)–\( t_{1/2} \) relation. By adding \( M_{14} \), the model reduces the scatter in the relative error to 0.251 while slightly increasing the median overestimation bias to 5.3\%, an increase of 0.7\% compared to using \( M_{12} \) alone. The final linear model is given by:  
\begin{align}
t_{1/2} \text{(Gyr)} = -3.53 M_{12} + 0.19 {M_{BCG}}/{M_{sat}} -1.48 M_{14} + 9.91 \label{equation:5}
\end{align}  
where the definitions of \( M_{12} \), \( M_{BCG}/M_{sat} \), and \( M_{14} \) are provided in Table \ref{Table:2}.  
Lastly, we have also verified that including the \textit{com\textunderscore offset} only marginally improves the linear models by reducing the scatter. This is because the \textit{com\textunderscore offset} is tightly correlated with $M_{12}$. For the sake of simplicity, we did not include this information in these linear models.

\section{Conclusions}\label{section:6}

In this paper, we developed machine learning models to predict $t_{1/2}$ using theoretical and observationally motivated properties derived from \thethree\ \textsc{Gizmo-Simba} simulations. From these simulations, we extracted 1,918 dark matter halos within a broad mass range of \(2.50 \times 10^{13} h^{-1} M_{\odot} - 2.64 \times 10^{15} h^{-1} M_{\odot}\) using AHF halo catalogs. The formation time for these halos, defined as the epoch at which a halo accretes half of its final mass, was computed using halo merger trees constructed with the MERGERTREE tool within AHF. 

Our random forest models demonstrated that $t_{1/2}$ predictions can achieve high accuracy by capturing non-linear relationships between halo formation time and various properties, including halo properties (Table \ref{Table:1}), BCG-ICL properties (Table \ref{Table:2}), and gas and stellar properties (Table \ref{Table:3}). Among the six models shown in Figure \ref{fig:3}, Model 6, which utilized all available features, achieved the best balance between bias and the lowest variability in relative error, emphasizing the importance of comprehensive property inclusion. The feature importance plots in Figure \ref{fig:5} highlighted properties such as \textit{com\textunderscore offset}, $M_{12}$, and followed by others as crucial for prediction accuracy. We refer the reader to Figure \ref{fig:5} for a complete description of the important features.

To focus on properties that are more observationally driven and directly aid in predicting \( t_{1/2} \), we trained CNNs using six baryonic properties: \( Z \), \( t_{\mathrm{age}} \), \( M \), \( M_{\mathrm{gas}} \), \( Z_{\mathrm{gas}} \), and \( T_{\mathrm{gas}} \). Using three different halo property map datasets—each based on a distinct binning strategy—we trained three separate CNN models with the architecture shown in Figure \ref{fig:7} to examine how binning methodology affects \( t_{1/2} \) predictions. The CNN models exhibited slightly lower bias in \( t_{1/2} \) predictions compared to random forest models; however, the scatter in relative prediction error was approximately 5\% higher (see Figure \ref{fig:8}). Based on the CNN results, we defined optimal radial ranges for each property (listed in Table \ref{Table:4} for all three binning approaches), which enabled the training of simplified random forest models. These targeted models showed slightly reduced accuracy compared to the CNNs but performed robustly, achieving biases of 7.3–9.7\% with manageable scatter in the relative prediction error, highlighting the robustness of CNN-informed feature selection. We further investigated in \ref{section:correlation and EPS} the performance of the RF and CNN models in reproducing key correlation trends, such as \(M_{200c}-t_{1/2}\) and \(c_{\mathrm{NFW}}-t_{1/2}\), and also assessed how the performance of the ML models compares to the analytical formalism of \cite{10.1093/mnras/stv689}.
The RF and CNN models successfully reproduce the key correlation trends between \(t_{1/2}\) and properties such as \(M_{200c}\) and \(c_{\mathrm{NFW}}\). 
Benchmarking against both simulation-based formation times and the analytical formalism shows that the models capture the underlying physical trends more closely to the simulation, i.e., the actual truth, and demonstrate that both RF and CNN models effectively learn to predict the formation time of the halos. 

Finally, we developed three simple linear models using key observable properties from Table \ref{Table:2} and Table \ref{Table:3}, achieving competitive accuracy while reducing the complexity of CNN and RF models. These linear models provide a practical framework for observers to estimate \( t_{1/2} \) directly. The baseline model, which uses only \( M_{12} \), has a median overestimation bias of 4.6\% and a scatter of 0.271. Adding observable properties such as the ratio of the total BCG stellar mass to the total satellite stellar mass within the halo (\( M_{BCG}/M_{sat} \)) and the magnitude/mass gap (\( M_{14} \)) further improves the scatter while maintaining accuracy. The linear model incorporating both \( M_{12} \) and \( M_{BCG}/M_{sat} \) reduces the scatter to 0.264 with a 3.5\% bias, while the model including \( M_{12} \), \( M_{BCG}/M_{sat} \), and \( M_{14} \) further decreases the scatter to 0.251 with a 5.3\% bias. We also provide the mathematical relations for these three linear models in Equations \ref{equation:3}, \ref{equation:4}, and \ref{equation:5}, allowing observers to directly estimate \( t_{1/2} \). Although these models slightly increase the median relative error bias in predicting \( t_{1/2} \), they strike a good balance by capturing essential information in a simpler form, reducing model complexity without sacrificing predictive quality.  
The practical application of these linear models depends largely on the observational context. Although \(t_{1/2}\) is a theoretical quantity derived from simulations and cannot be directly measured in observations, these relations can serve as useful approximations by employing observable proxies. For instance, in optical surveys such as those conducted with HST, quantities like the magnitude gaps \(M_{12}\) and \(M_{14}\), as well as the mass ratio \(M_{\mathrm{BCG}}/M_{\mathrm{sat}}\), are relatively straightforward to determine—though some level of scatter in the inferred \(t_{1/2}\) is expected. Interestingly, the \(f_{\mathrm{BCG+ICL}}\) fraction, which is suggested to correlate strongly with halo formation time by \cite{kimmig2025intra}, does not pop up in our studies, indicated as $f_{*, 100}$. Though there are some differences to the \(f_{\mathrm{BCG+ICL}}\) fraction, and it is the most important feature in our Model 3 (see \autoref{fig:5}), compared to other quantities, $f_{*, 100}$ is much less significant. One simple explanation is that this quantity is strongly correlated with the other important features. This is also supported by both observation \citep{2018ApJ...857...79J} and simulation \citep{contreras2024characterising} results -- the dynamical relaxed clusters tend to have lower ICL fraction compared to these un-relaxed clusters.

This study highlights the potential of combining machine learning to refine halo property predictions, reducing model complexity while preserving accuracy. However, the results are specific to simulations using Gizmo-Simba physics. Future work will explore the impact of varying baryonic physics on the prediction performance.

\section*{Acknowledgements}

AS and WC are supported by the Atracci\'{o}n de Talento Contract no. 2020-T1/TIC-19882 granted by the Comunidad de Madrid in Spain. WC also thanks the Agencia Estatal de Investigación (AEI) for the Consolidación Investigadora Grant CNS2024-154838, the science research grants from the China Manned Space Project and HORIZON EUROPE Marie Sklodowska-Curie Actions for supporting the LACEGAL-III project with grant number 101086388. WC and DdA thank the Ministerio de Ciencia e Innovación (Spain) for financial support under Project grant PID2021-122603NB-C21. ER acknowledges support from the Chandra Theory Program (TM4-25006X) awarded from the Chandra X-ray Center which is operated by the Smithsonian Astrophysical Observatory for and on behalf of NASA under contract NAS8-03060. YZ is supported by the National Key Basic
Research and Development Program of China (No. 2023YFA1607800,
2023YFA1607804), the National Science Foundation of China (12173024), the China Manned Space Project (No. CMS-CSST-2021-A01, CMS-CSST-2021-A02, CMS-CSST-2021-B01), the ``111'' project of the Ministry of Education under grant No. B20019, and the generous sponsorship from Yangyang Development Fund. The work presented here emerged out of the annual The300 workshop held at UAM's La Cristalera during the week July 8-12, 2024, partially funded by the 'Ayuda para la Organización de Jornadas Científicas en la UAM en el Marco del Programa Propio de Investigación y con el Apoyo del Consejo Social de la UAM'

The authors acknowledge The Red Espa\~nola de Supercomputaci\'on for granting computing time for running the hydrodynamic simulations of \thethree\ galaxy cluster project in the Marenostrum supercomputer at the Barcelona Supercomputing Center.
\bibliographystyle{aa} 
\bibliography{ref} 

\begin{appendix}
\section{Histograms of the Maximum Bin Edges for the Halo Sample}\label{Appendix:1}
\begin{figure}[h]
    \centering
    \includegraphics[width=0.5\linewidth]{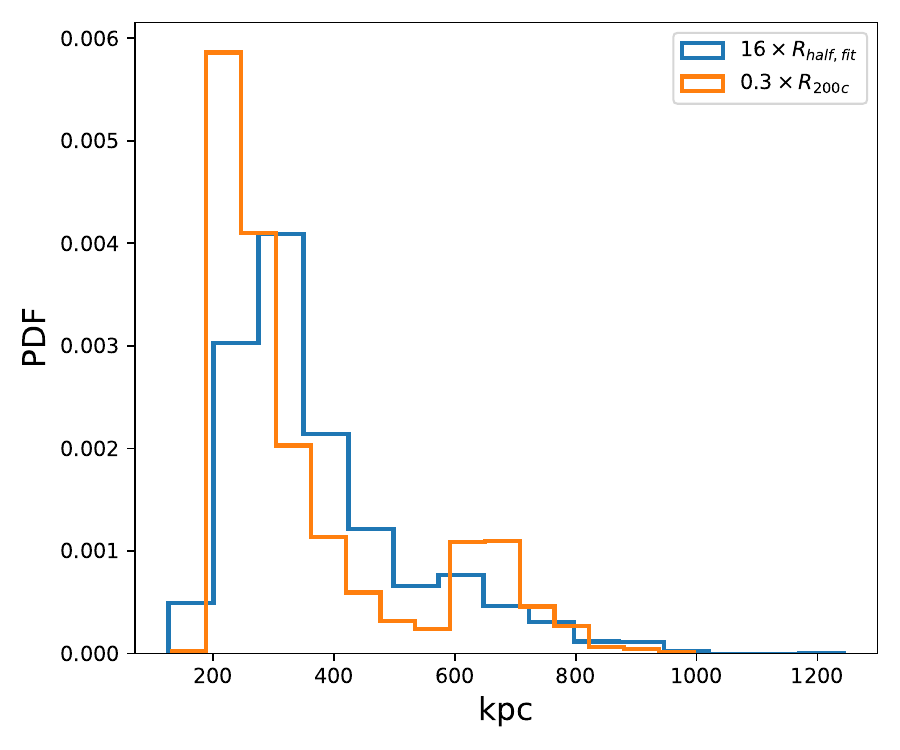}
    \caption{Histograms showing the distribution of $0.3 R_{200c}$ and $16 R_{\mathrm{half,fit}}$ for the halos used in this study.}
    \label{fig:A.1}
\end{figure}
In Figure \ref{fig:A.1}, we show the distributions of $0.3 \times R_{200c}$ and $16 \times R_{\mathrm{half,fit}}$ for the halos, representing the maximum edges of the bins used to construct property maps with Binning Methodology 2 and Binning Methodology 3.

\section{Saliency Maps}\label{Appendix:saliency}
\subsection{Interpreting CNN Performance and Identifying Key Property Aperture Ranges}
Understanding the results of CNNs and other deep learning models is challenging due to their high complexity, which stems from billions of interconnected parameters within the network. These parameters exhibit complex, non-linear relationships with each other, making it difficult to understand how each parameter influences the final output. This complexity further complicates the process of deriving clear insights from their outputs. Moreover, the network employs a series of activation functions from one layer to the next, which are not straightforward to interpret, and thus hinder the ability to trace the path from input to final output in the network. That is why CNN models are popularly known as 'black boxes' due to these complexities. Many methods are currently being devised to enhance the interpretability of CNN networks. In this subsection, we attempt to interpret our single-channel CNN network using the pixel attribution methodology discussed in \citet{molnar2020interpretable} to understand the contribution of each pixel in the input image to the prediction. We achieve this by employing the Vanilla Gradient-based approach introduced in \citet{simonyan2013deep}. This method works by forward propagating the image of interest through the network, then computing the pixel-wise gradient of the loss to the input image using the back-propagation algorithm, where the pixel values indicate how sensitive the model's output is to changes in the input features.

\begin{figure}[h]
    \centering
    \includegraphics[scale=0.5]{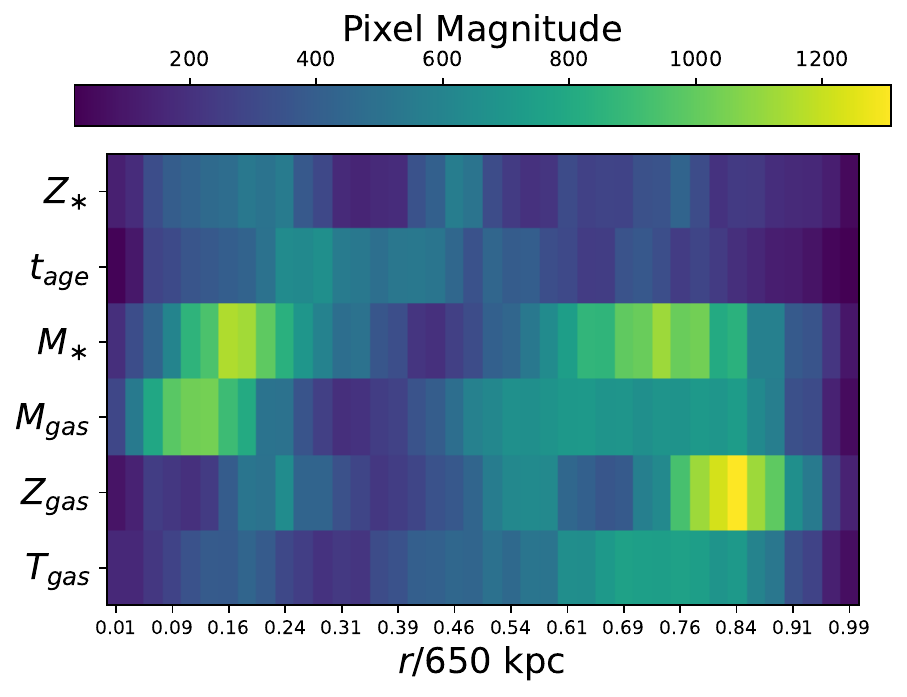}

    \includegraphics[scale=0.4]{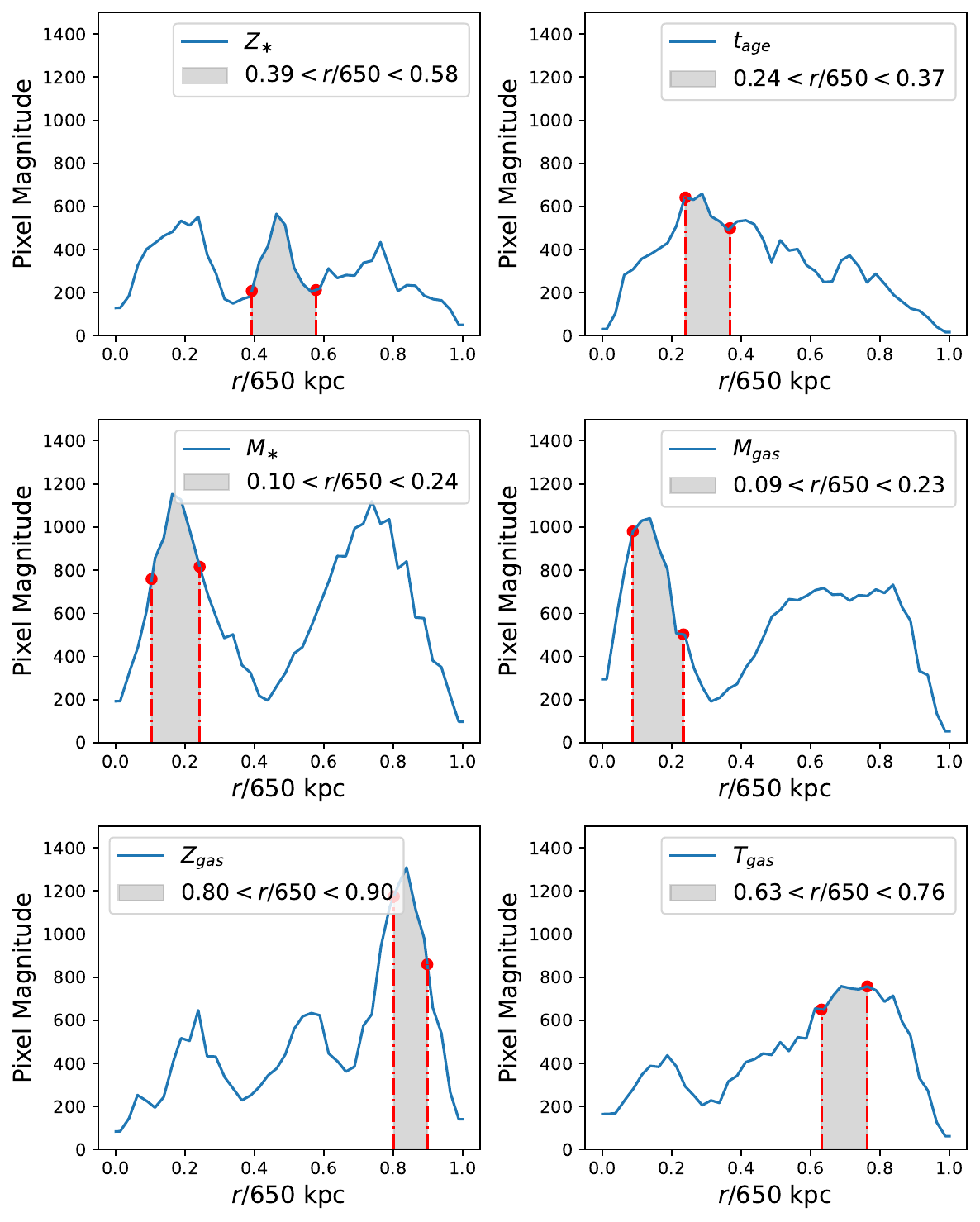}
    \caption{The image in the upper panel depicts the stacked saliency map for CNN Model 1, where the halo's central galaxy baryonic particles are partitioned into 40 uniform bins placed from the center of the halo out to 650 kpcs to generate input maps for training and testing.
    The six line plots in the lower panel show the pixel intensity variation with $r/650$ for each row of the stacked saliency map, where each row corresponds to a different physical property. The gray region in each plot denotes the key aperture range for the physical property that the network considers the most significant for making predictions. In each plot, the gray region covers approximately 22\% of the total area under the curve.}
    \label{fig:B.1}
\end{figure}

\begin{figure}[h]
    \centering
    \includegraphics[scale=0.5]{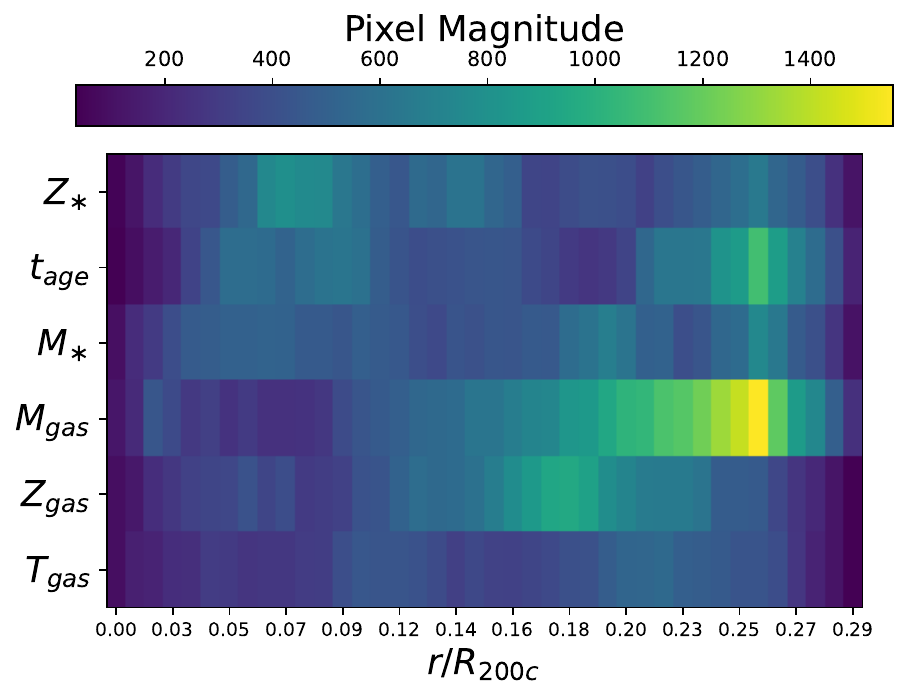}

    \includegraphics[scale=0.4]{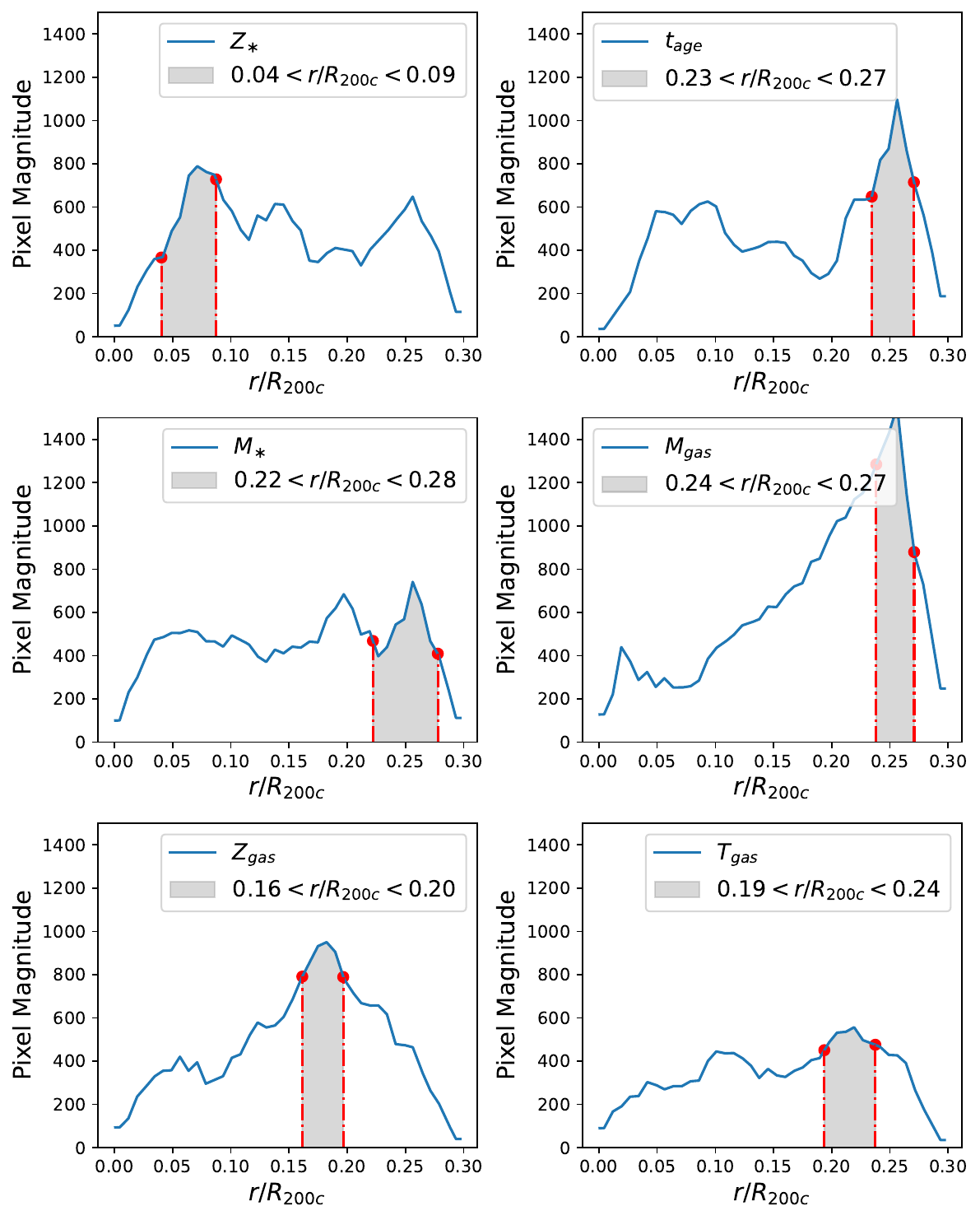}
    \caption{ A plot similar to Figure \ref{fig:B.1}, but for CNN Model 2, where the halo's central galaxy baryonic particles are partitioned into 40 uniform bins, ranging from the center of the halo out to 0.3 $R_{200c}$, to generate input maps used for training.
    }
    \label{fig:B.2}
\end{figure}

\begin{figure}[h]
    \centering
    \includegraphics[scale=0.5]{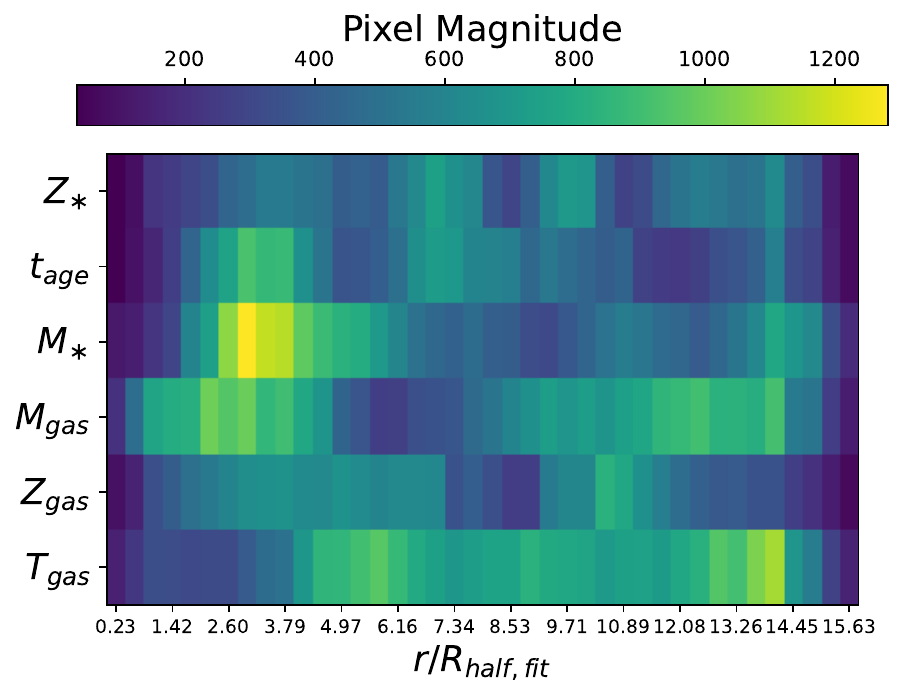}

    \includegraphics[scale=0.4]{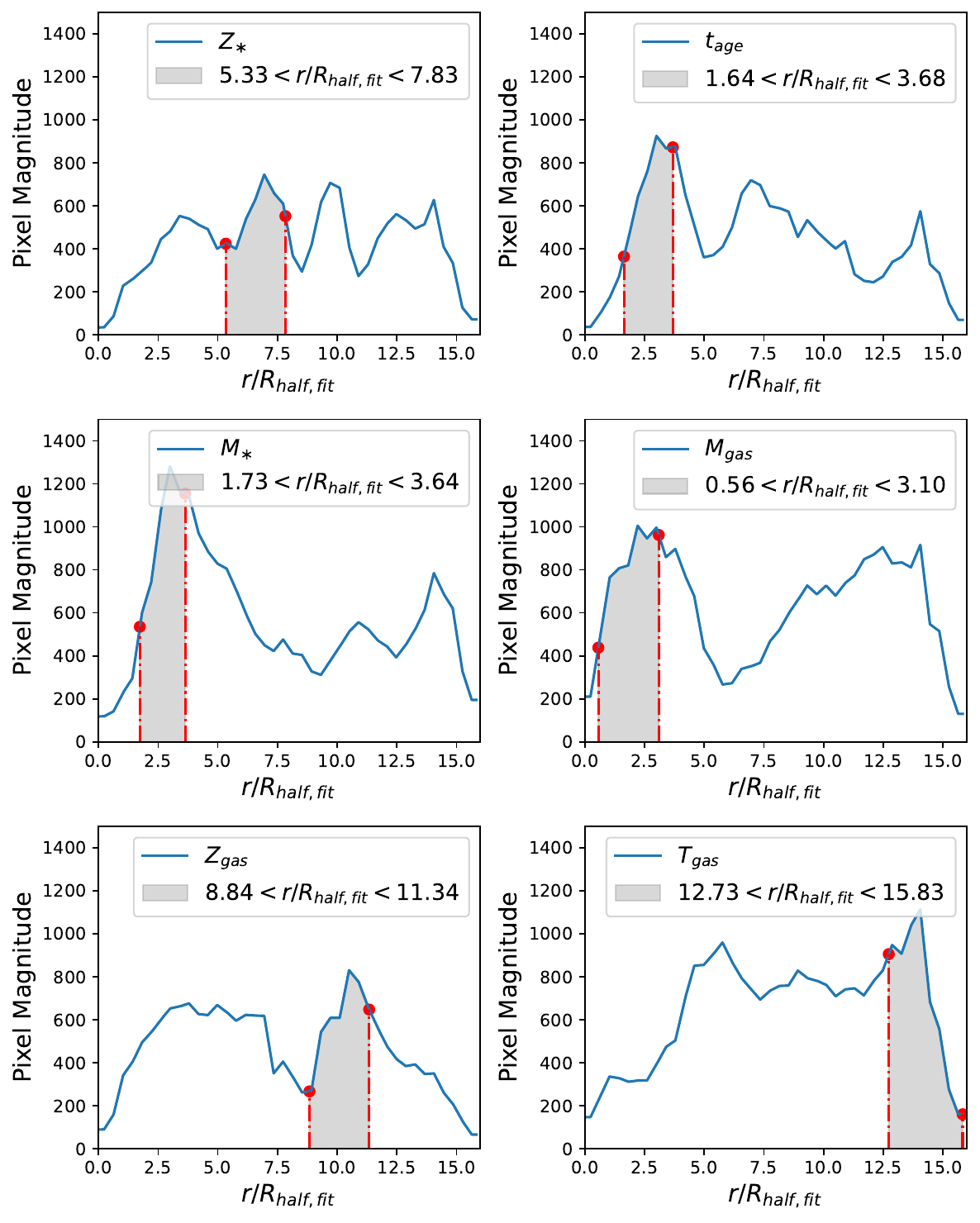}
    \caption{A plot similar to Figure \ref{fig:B.1}, but for CNN Model 3, where the halo's central galaxy baryonic particles are partitioned into 40 uniform bins, ranging from the center of the halo out to 16 $R_{\mathrm{half,fit}}$, to generate input maps used for training.
    }
    \label{fig:B.3}
\end{figure}

During the CNN training, we back-propagated each property map from the training and test set through the network to generate the saliency map with dimensionality the same as the input property map. The pixel values in each saliency map were then normalized between 0 and 1. The saliency maps for each halo were then stacked together to form the final map that contains statistical information on the region that significantly impacts the network prediction. In the upper panel of Figures \ref{fig:B.1}, \ref{fig:B.2}, and \ref{fig:B.3}, we present the final stacked saliency maps obtained from CNN Model 1, CNN Model 2, and CNN Model 3, respectively. The saliency maps obtained from the three CNN models highlight how variations in the training input maps, resulting from different binning criteria, affect the network's interpretability. It is clear that these highlighted regions as indicated by the color bar on top have a higher effect on the CNN's predictions: $M_{*}$, $M_{\mathrm{gas}}$ and $Z_{\mathrm{gas}}$ for model 1; $M_{\mathrm{gas}}$ and $Z_{\mathrm{gas}}$ for Model 2; $M_{*}$, $M_{\mathrm{gas}}$ and $T_{\mathrm{gas}}$ for model 3.
The significance of stellar mass in predicting the assembly history of the halo has been suggested in various studies, such as \citet{,yang2006observational,lin2016detecting, lim2016observational}. However, the saliency maps from different models give different predictions, not only the properties but also their positions, which are detailed in the following paragraph.

In the lower panels of Figures \ref{fig:B.1}, \ref{fig:B.2}, and \ref{fig:B.3}, we 
show the profiles of the saliency maps for all six properties.
To determine the most significant aperture range for each CNN model/binning methodology, which will be used later for a simpler model to predict the $t_{1/2}$, we identified the position where the pixel value is at its maximum. From this maximum position, we expanded the area by 10\%
of the total area under the curve above and below this position to obtain the upper and lower radial bounds corresponding to each property in the saliency map. Slightly increasing or decreasing this range has a very minor effect on the model's prediction power. 
In the lower panel of Figure \ref{fig:B.1}, the identified aperture ranges for the six properties of CNN Model 1 are shown as shaded grey regions, and these ranges are listed in the first column of Table \ref{Table:4}. From the distribution plots of CNN Model 1, we can see that 
identifying the regions to consider is not as straightforward. 
For example, the $M_{*}$ distribution shows two nearly identical peaks, indicating that both regions 
are equally important for this property. The same can be stated for $Z_{*}$, where we find nearly three identical peaks.
The aperture ranges for the six properties of CNN Model 2, as identified in the lower panel of Figure \ref{fig:B.2}, are listed in the second column of Table \ref{Table:4}. Note that the most important property for CNN Model 2 is gas mass. The halos vary significantly in size, with more massive halos having larger \(R_{200c}\). This differs from binning based on a fixed distance of 650 kpc, which may not be able to fully probe the interior regions of massive halos when the same bin is applied at different positions for halos of varying mass. Model 2 appears to yield more consistent results, as quantities in bins that scale with \(R_{200c}\) are on equal footing for halos of different sizes, meaning the quantities are estimated relative to each halo's size. However, such estimations rely on accurately measuring \(R_{200c}\), which is not straightforward to determine from observations.
Finally, the aperture ranges for the six properties for model 3 are shown in the lower panel of Figure \ref{fig:B.3}, with their radii ranges listed in the third column of Table \ref{Table:4}. 
The most important property in this case is the stellar mass. Similar to the saliency map for binning Method 1, also this map is noisier compared to the saliency map of Method 2 (i.e. Figure \ref{fig:B.2}). 

The aperture ranges for almost all the properties, as estimated using a stacked saliency map for the three different CNN models, indicate that the region the network deems most important lies well beyond the BCG region, plus the area beyond the transitional radius (i.e., the distance where the ICL starts to dominate the BCG stellar component), especially for CNN Model 1.
Various literature sources have reported estimates for this transition radius, which vary depending on the methodologies employed to identify the ICL. Observational studies have found this range to be 60-80 kpc \citep{montes2021buildup}, with slightly increased values reported by \citet{zhang2019dark} at 100 kpc, and \citet{chen2022sphere} at 70-200 kpc. Conversely, \citet{contini2021brightest, contini2021origin, contini2022transition} used numerical simulations to show that the transition radius is $60 \pm 40$ kpc. In our study, this best radius range for predicting the $t_{1/2}$  
depends on the binning method as well as on the baryonic properties, i.e. the proposed transition radius from observations may not be the best indicator for $t_{1/2}$.

\begin{table*}[h]
    \renewcommand{\arraystretch}{1.5}     \centering
    \caption{Selected radius ranges for the six properties identified using the stacked saliency maps for the three trained CNN models.
    The feature importance ranking in the table corresponds to the RF models shown in Figure \ref{fig:B.4}, which are trained directly on the property calculated within the saliency ranges. 
    }
    \label{Table:4}
    \begin{tabular}{|p{1.5cm}|p{1.7cm}|p{1.7cm}|p{1.7cm}|p{1.7cm}|p{1.7cm}|p{1.7cm}|}
        \hline
 \textbf{Property} & \multicolumn{2}{c|}{\textbf{CNN Model 1}} & \multicolumn{2}{c|}{\textbf{CNN Model 2}} & \multicolumn{2}{c|}{\textbf{CNN Model 3}} \\
        \cline{2-7}
        & \textbf{Range} \newline $\times$ 650 kpc & \textbf{Feature} \newline \textbf{Importance} 
        & \textbf{Range} \newline $\times$ $R_{200c}$ & \textbf{Feature} \newline \textbf{Importance} 
        & \textbf{Range} \newline $\times$ $R_{\mathrm{half,fit}}$ & \textbf{Feature} \newline \textbf{Importance} \\
        \hline
        $Z_{*}$ & 0.39-0.58 & 5 & 0.04-0.09 & 6 & 5.33-7.83 & 6\\
        $t_{\mathrm{age}}$ & 0.24-0.37 & 2 & 0.23-0.27 & 1 & 1.64-3.68 & 3 \\
        $M_{*}$ & 0.10-0.24 & 3 & 0.22-0.28 & 3 & 1.73-3.64 & 2\\
        $M_{\mathrm{gas}}$ & 0.09-0.23 & 4 & 0.24-0.27 & 2 & 0.56-3.10 & 4 \\
        $Z_{\mathrm{gas}}$ & 0.80-0.90 & 6 & 0.16-0.20 & 5 & 8.84-11.43 & 5 \\
        $T_{\mathrm{gas}}$ & 0.63-0.76 & 1 & 0.19-0.24 & 4 & 12.73-15.83 & 1 \\
        \hline
    \end{tabular}
\end{table*}

\subsection{Random forest model using properties calculated within aperture ranges}
We used the saliency ranges of the properties listed in Table \ref{Table:4} from the three CNN models, which were trained on input maps generated using three distinct binning methodologies, to train three new random forest models. This task aims (1) to verify that the selected regions can provide similar predictions as the CNN models, i.e., that these are the most important regions for predicting $t_{1/2}$, and (2) to develop a machine learning model that is less complex than CNN or RF models utilizing an extensive list of properties to enable a more targeted analysis.  By restricting the input properties to specific radial ranges, our aim is to check the influence of different spatial regions on halo formation time predictions, providing a clearer physical interpretation. This approach may make inputs less directly comparable to observational data, but it aligns the way observational investigations focus on using various apertures rather than the whole global properties. The CNN trained on galaxy properties was used to infer the radial ranges that are most informative for predicting halo formation time. Retraining the random forest models with these targeted inputs will help us to assess how the identified radial regions contribute to predictive power, compared to a model trained on all available simulation data.  While this does introduce additional training steps, it allows us to explicitly test whether selecting specific radial regions improves predictions in a way that could be useful for observational studies. 
These three new random forest models will be trained directly using the stellar and gas properties within the saliency range. Again, the training and test sets comprise the same 1,630 and 288 halos used in Section \ref{section:3} and for the CNN models discussed previously.
For each halo in the training or test set, we computed \(Z_{*}\), \(t_{\mathrm{age}}\), \(M_{*}\), \(M_{\mathrm{gas}}\), \(Z_{\mathrm{gas}}\), and \(T_{\mathrm{gas}}\) for the aperture ranges given in Table \ref{Table:4} for the three distinct saliency maps.
In this approach, we use fewer properties or features than the random forest models discussed in Section \ref{section:3}, focusing on baryonic properties that are now easier to obtain from observations. The aperture radius for all the properties is determined based on the saliency maps, which highlight the regions within the input maps that the network considers most important for making predictions. This approach is more robust because it allows for an informed decision regarding the aperture range to be used for training the machine learning models, as opposed to the arbitrary criteria we used previously, where we simply considered aperture ranges of 30 kpc, 50 kpc, and 100 kpc. 
The random forest training setup is the same as in section \ref{section:3}. We used the same hyperparameters and their value ranges as those used to train the six different random forest models in section \ref{section:3}. Additionally, we used the same sampling weights for the training set to assign weights to each training data to handle the undersampling issue.
 
\begin{figure}[h]
    \centering
    \includegraphics[scale=0.41]{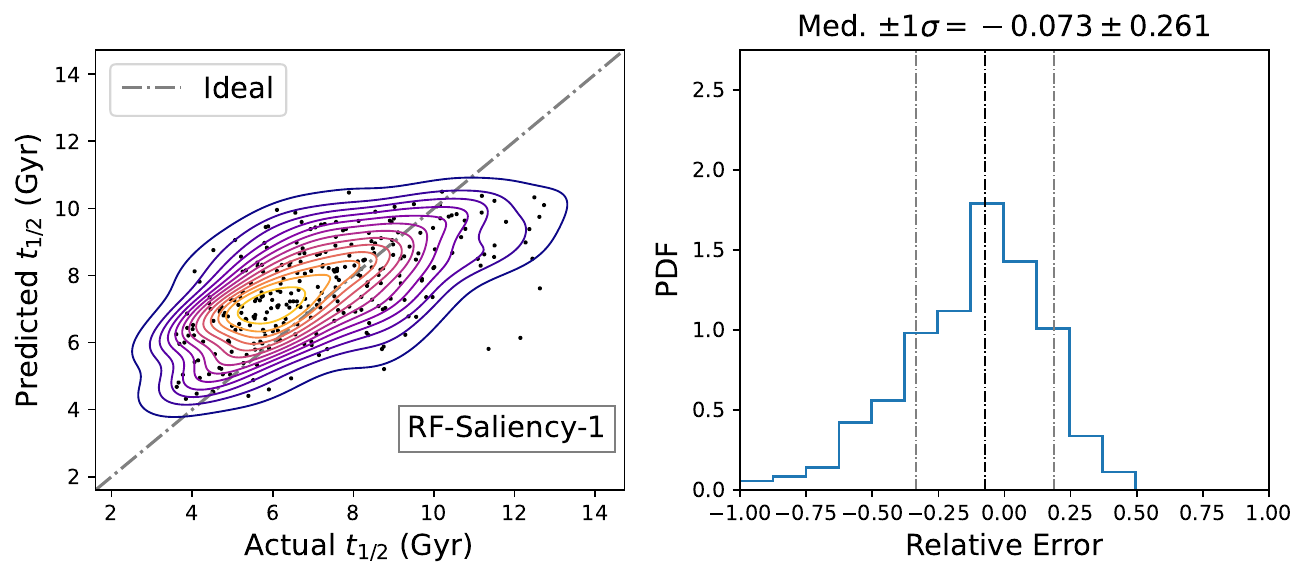}

    \includegraphics[scale=0.41]{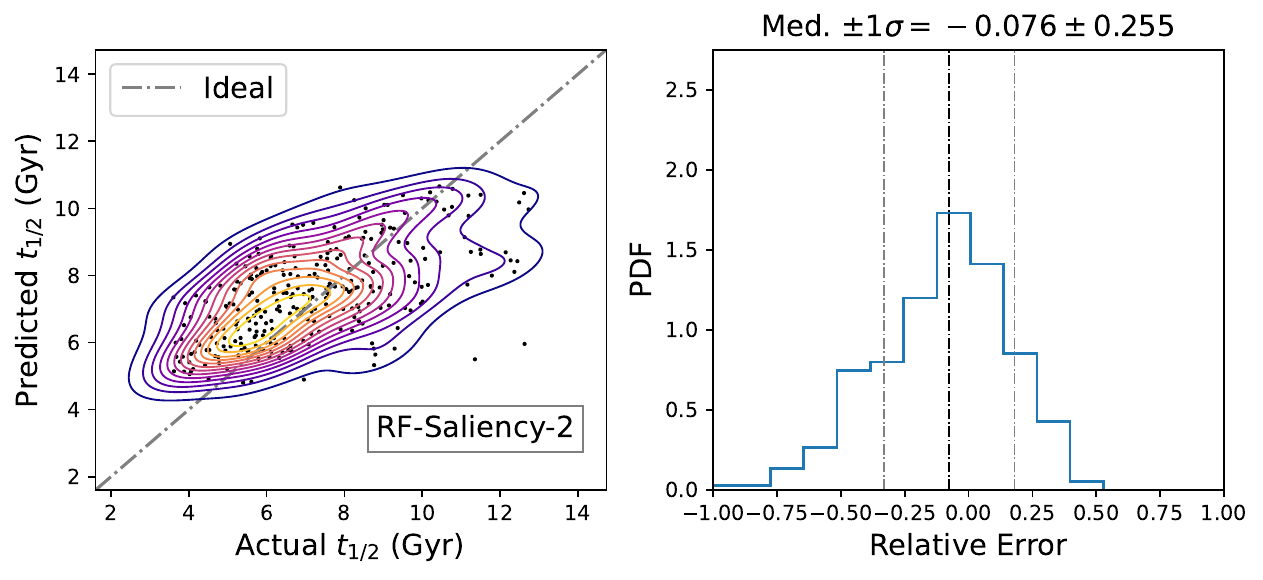}

    \includegraphics[scale=0.41]{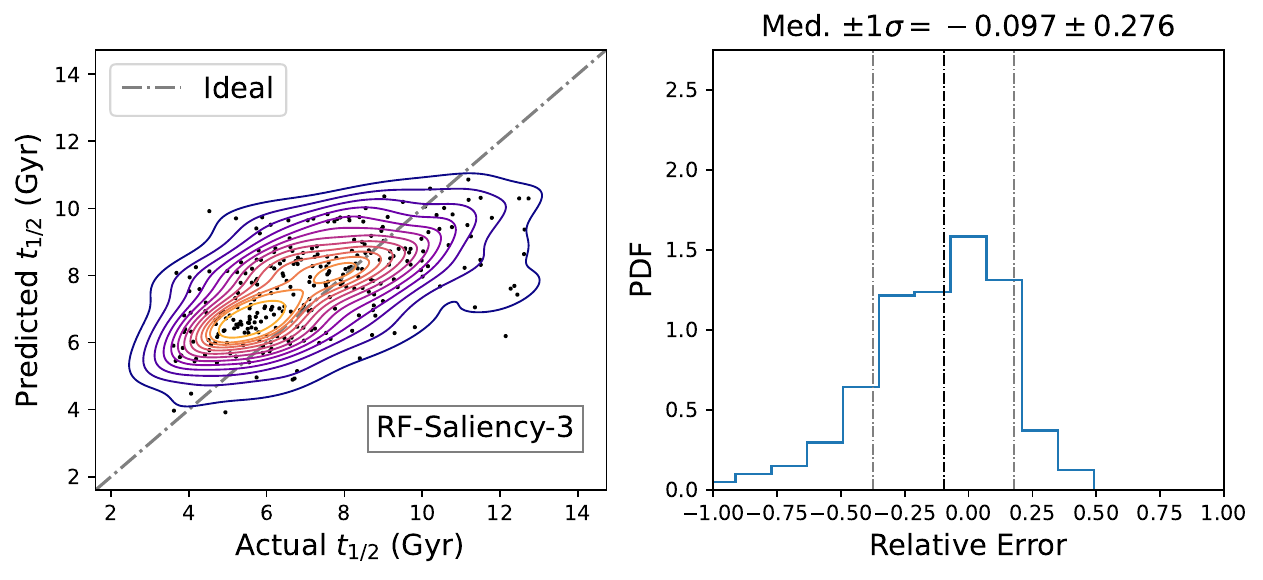}
    \caption{
    This figure compares the predicted $t_{1/2}$ values for the three random forest models with their actual values from the test dataset. Each model is trained on properties derived from the saliency map-identified aperture radial range all listed in Table \ref{Table:4}. The first panel displays results for the random forest model trained with the aperture range from CNN Model 1, as detailed in the first column of Table \ref{Table:4}. The second panel presents results for the model trained with the aperture range from CNN Model 2, also listed in the first column of Table \ref{Table:4}. The third panel shows results for the model trained with the aperture range from CNN Model 3, with aperture cuts listed in the third column of Table \ref{Table:4}.
    Each plot in the left column displays a 2D joint probability density function comparing the true formation time values from the test dataset with the formation time values predicted by the random forest model that utilizes the saliency range. Black data points represent the predicted $t_{1/2}$ values from this new random forest model against the true $t_{1/2}$ values, illustrating their spread relative to the ideal model, which is depicted by the 45-degree dotted grey diagonal line. Each plot in the right column displays the probability density function of the relative errors between the true formation time values and the predictions from the random forest model, along with the median error and the standard deviation.
    }
    \label{fig:B.4}
\end{figure}

In Figure \ref{fig:B.4}, we present the $t_{1/2}$ predictions for the test set along with the actual values and the histograms of the relative errors for all three trained random forest models. The test results for the random forest model shown in the upper panel of Figure \ref{fig:B.4} (i.e. RF-Saliency-1) correspond to the model where properties were computed using saliency ranges identified from the saliency map of CNN model 1  (i.e. first column of Table \ref{Table:4}). Similarly, the middle (i.e. RF-Saliency-2) and lower (i.e. RF-Saliency-3) panels display results for the random forest models trained using properties within the saliency ranges (i.e. second and third column of Table \ref{Table:4}) identified from the saliency maps for CNN Model 2 and CNN Model 3, respectively.

Upon comparing the CNN models (Figure \ref{fig:8}) with these random forest models trained using properties within the saliency range (Figure \ref{fig:B.4}, the results show that the prediction accuracy of the random forest models degrades slightly when using properties estimated within the saliency range for training.
In the scatter and contour density plots, the CNN models' predictions are more closely clustered around the ideal line, indicating more accurate predictions, while the RF models exhibit a slightly larger deviation from the ideal model line.
This is because the median of the relative error increased slightly by a few per cent when we directly used the saliency range properties and also fewer features compared to the three CNN models. The median relative error bias is lower for CNN models (4.1\% for Model 1, 1.6\% for Model 2, and 0.4\% for Model 3) than for RF models (7.1\% for RF Saliency-1, 7.6\% for RF Saliency-2, and 9.7\% for RF Saliency-3), suggesting that CNN predictions are closer to the ideal model line, and the actual $t_{1/2}$ values. The spread or the standard deviation of the relative errors also increased by approximately 1 to 3 per cent when we directly used saliency range properties compared to the CNN models that were trained using input maps that are slightly more complex and consist of more features. The standard deviation of the relative error histogram increased from 0.251 for CNN Model 1 to 0.26 for RF Saliency-1, and from 0.233 for CNN Model 2 to 0.255 for RF Saliency-2, while remaining the same for CNN Model 3 and RF Saliency-3 when we compared CNN models prediction quality with the random forest model that relied on training with properties within the saliency range. 

In this section, we demonstrate that using only the gas and stellar properties within the salient aperture range achieves a performance comparable to that of training the full CNN model with the entire property maps, as well as the six random forest models discussed in Section \ref{section:3}, which utilize a diverse set of features. This is a very promising result because these predictions do not involve any satellite galaxies, which will be combined in the next section to provide the best linear model for predicting $t_{1/2}$. In Table \ref{Table:4}, we provide the feature importance ranking for all six properties across all three RF models.

To assess the reliability of the saliency maps in identifying important regions for predicting \(t_{1/2}\), we conducted a series of targeted analyses. We examined first the impact of selecting alternative regions—both around dominant peaks and valleys—from the saliency map profiles of key baryonic properties, say \(M_{*}\) in the lower panel of Figure \ref{fig:B.1} or \(M_{gas}\) as given in the lower panel of Figure \ref{fig:B.2}. Retraining the Random Forest model using these regions resulted in only minor variations in predictive performance, suggesting a limited sensitivity to the precise choice of radial interval. We further investigated by computing the Pearson correlation coefficients between the baryonic properties and  \(t_{1/2}\) within each bin, focusing on both raw and transformed properties. For CNN Model 2 (with \(R_{200c}\)-scaled bins), we observed that while \(M_{gas}\) consistently appeared as the most correlated feature, the overall correlation remained low (typically < 0.35) and relatively flat across bins. Similar trends were noted for CNN Model 1 (fixed 650 kpc binning). These findings suggest that the saliency maps, although somewhat aligned with correlation plots, they are noisy and do not strongly correlate with actual physical trends in the data. As a result, we conclude that while saliency maps can offer a qualitative sense of which features the model considers important, they should be interpreted with caution—particularly because the underlying baryonic properties show weak statistical correlation with \( t_{1/2} \).
\end{appendix}
\end{document}